\documentclass[11pt]{article}

\usepackage[utf8]{inputenc}
\usepackage[T1]{fontenc}

\usepackage{bm}
\usepackage{dsfont}
\usepackage{csquotes}

\usepackage{xspace}
\usepackage{nicefrac}
\usepackage[super]{nth}
\usepackage{microtype}

\usepackage[english]{babel}

\usepackage[shortlabels]{enumitem}

\usepackage{booktabs}
\usepackage{makecell}

\usepackage{graphicx}
\usepackage{subcaption}

\usepackage{xcolor}

\usepackage{indentfirst}

\usepackage{algorithm}
\usepackage[noend]{algpseudocode}
\algrenewcommand{\algorithmiccomment}[1]{$\vartriangleright$ #1}
\algrenewcommand{\algorithmicreturn}{\textbf{Return: }}
\algnewcommand\algorithmicinput{\textbf{Input: }}
\algnewcommand\Input{\State \algorithmicinput}

\usepackage{titletoc}

\usepackage[colorlinks=true,bookmarks=true,linkcolor=blue,
urlcolor=blue,citecolor=blue,breaklinks=true]{hyperref}

\usepackage{url}

\makeatletter
\addto\extrasenglish{%
\renewcommand{\sectionautorefname}{\S\@gobble}%
\renewcommand{\subsectionautorefname}{\S\@gobble}%
\renewcommand{\subsubsectionautorefname}{\S\@gobble}%
}
\makeatother

\usepackage{changepage}                 
\newlength{\offsetpage}
{\end{adjustwidth}}

\usepackage{amsmath}
\usepackage{amsfonts} 
\usepackage{amssymb}
\usepackage{mathrsfs}
\usepackage{mathtools} 
\mathtoolsset{showonlyrefs}

\usepackage{pgfplots}

\usepackage[noend]{algpseudocode}

\usepackage{footnote} 
\makesavenoteenv{algorithm}

\usepackage{amsthm}

\newtheorem{theorem}{Theorem}

\newtheorem{definition}{Definition}

\renewcommand{\phi}{\varphi}

\renewcommand{\leq}{\leqslant}
\renewcommand{\geq}{\geqslant}

\renewcommand{\epsilon}{\varepsilon}
\renewcommand{\imath}{\mathrm{i}}

\usepackage{accents}

\newlength{\restsubwidth}
\newlength{\restsubheight}
\newlength{\restsubmoreheight}
\newcommand{\rest}[2]{%
        \settowidth{\restsubwidth}{\ensuremath{#2}}
        \settoheight{\restsubheight}{\ensuremath{{}_{#2}}}
        \ensuremath{{#1\hskip 0.5pt}_{\vrule\kern2pt\parbox[b][%
        4pt][b]{\the\restsubwidth}{%
                        \ensuremath{{}_{#2}}}}}
        }

\graphicspath{{./images/}}
\usepackage[round]{natbib}

\usepackage{booktabs}
\usepackage{breakcites}
\usepackage{authblk}

\usepackage[margin=1in]{geometry}

\title{A Permutation-Equivariant Neural Network Architecture For Auction Design}

\author[a]{Jad Rahme \thanks{Corresponding author: \href{jrahme@princeton.edu}{jrahme@princeton.edu}  }}
\author[a,b]{Samy Jelassi}
\author[b, c, d]{Joan Bruna}
\author[a]{S. Matthew Weinberg}
\affil[a]{Princeton University, Princeton}
\affil[b]{Courant Institute of Mathematical Sciences, New York University, New York}
\affil[c]{Center for Data Science, New York University, New York}
\affil[d]{Institute for Advanced Study, Princeton}

\makeatletter
\def\imod#1{\allowbreak\mkern10mu({\operator@font mod}\,\,#1)}
\makeatother

\makeatletter
\def\inmod#1{\allowbreak\mkern5mu({\operator@font mod}\,\,#1)}
\makeatother
\usepackage{savesym}
\usepackage{xcolor}
\usepackage{amsthm}
\usepackage{amsmath}
\usepackage{amssymb}

\usepackage{hyperref}

\hypersetup{
     colorlinks = true,
     linkcolor = brown,
     anchorcolor = blue,
     citecolor = teal,
     filecolor = blue,
     urlcolor = blue
     }

\pgfplotsset{compat=1.14}

\begin{document}

\maketitle

\begin{abstract}
  Designing an incentive compatible auction that maximizes expected revenue is a central problem in Auction Design. Theoretical approaches to the problem have hit some limits in the past decades and analytical solutions are known for only a few simple settings. Computational approaches to the problem through the use of LPs  have their own set of limitations. Building on the success of deep learning, a new approach was recently proposed by \cite{dutting2017optimal} in which the auction is modeled by a feed-forward neural network and the design problem is framed as a learning problem. The neural architectures used in that work are general purpose and do not take advantage of any of the symmetries the problem could present, such as permutation equivariance. In this work, we consider auction design problems that have permutation-equivariant symmetry and construct a neural architecture that is capable of perfectly recovering the permutation-equivariant optimal mechanism, which we show is not possible with the previous architecture. We demonstrate that permutation-equivariant architectures are not only capable of recovering previous results, they also have better generalization properties.
  
\end{abstract}

\section{Introduction}


Designing truthful auctions is one of the core problems that arise in economics. Concrete examples of auctions include sales of treasury bills, art sales by Christie’s or Google Ads. Following seminal work of Vickrey~\citep{vickrey1961counterspeculation} and Myerson~\citep{myerson1981optimal}, auctions are typically studied in the \emph{independent private valuations} model: each bidder has a valuation function over items, and their payoff depends only on the items they receive. Moreover, the auctioneer knows aggregate information about the population that each bidder comes from, modeled as a distribution over valuation functions, but does not know precisely each bidder's valuation. Auction design is challenging since the valuations are private and bidders need to be encouraged to report their valuations truthfully. The auctioneer aims at designing an incentive compatible auction that maximizes revenue.   

While auction design has existed as a subfield of economic theory for several decades, complete characterizations of the optimal auction only exist for a few settings. Myerson resolved the optimal auction design problem when there is a single item for sale \citep{myerson1981optimal}.  However, the problem is not completely understood even in the extremely simple setting with just a single bidder and two items. While there have been some partial characterizations~\citep{manelli2006bundling, manelli2010bayesian,Pavlov11, WangT14, daskalakis2017strong}, and algorithmic solutions with provable guarantees~\citep{Alaei11, AlaeiFHHM12, AlaeiFHH13, CaiDW12a, CaiDW12b}, neither the analytic nor algorithmic approach currently appears tractable for seemingly small instances.


 Another line of work to confront this theoretical hurdle consists in building automated methods to find the optimal auction. Early works \citep{conitzer2002complexity,conitzer2004self} framed the problem as a linear program. However, this approach suffers from severe scalablility issues as the number of constraints and variables is exponential in the number of bidders and items \citep{guo2010computationally}. Later, \citet{sandholm2015automated} designed algorithms to find the optimal auction. While scalable, they are however limited to specific classes of auctions known to be incentive compatible. A more recent research direction consists in building deep learning architectures that design auctions from samples of bidder valuations. \citet{dutting2017optimal} proposed RegretNet, a feed-forward architecture to find near-optimal results in several known multi-item settings and obtain new mechanisms in unknown cases. This architecture however is not data efficient and can require a large number of valuation samples to learn an optimal auction in some cases. This inefficiency is not specific to RegretNet but is characteristic of neural network architectures that do not incorporate any inductive bias.
 
 %
 In this paper, we build a deep learning architecture for \textit{multi-bidder symmetric auctions}. These are auctions which are invariant to relabeling the items or bidders. More specifically, such auctions are \emph{anonymous} (in that they can be executed without any information about the bidders, or labeling them) and \emph{item-symmetric} (in that it only matters what bids are made for an item, and not its a priori label). 
 
 It is now well-known that when bidders come from the same population that the optimal auction itself is anonymous. Similarly, if items are \emph{a priori indistinguishable} (e.g. different colors of the same car --- individuals certainly value a red vs.~blue car differently, but there is nothing objectively more/less valuable about a red vs.~blue car), the optimal auction is itself item-symmetric. In such settings, our approach will approach the true optimum 
in a way which retains this structure (see Contributions below). Even without these conditions, the optimal auction is often symmetric anyway: for example, ``bundling together'' (the auction which allows bidders to pay a fixed price for all items, or receive nothing) is item-symmetric, and is often optimal even when the items are a priori distinguishable. 
 
Beyond their frequent optimality, such auctions are desirable objects of study \emph{even when they are suboptimal}. For example, seminal work of Hartline and Roughgarden which pioneered the study of ``simple vs.~optimal auctions'' analyzes the approximation guarantees achievable by anonymous auctions~\cite{HartlineR09}, and exciting recent work continues to improve these guarantees~\cite{AlaeiHNPY15,JinLTX19,JinLQTX19}. Similarly, \citet{daskalakis2012symmetries} develop algorithms for item-symmetric instances, and exciting recent work show how to leverage item-symmetric to achieve near-optimal auctions in completely general settings~\cite{KothariMSSW19}. To summarize: symmetric auctions are known to be optimal in many settings of interest (even those which are not themselves symmetric). Even in settings where they are not optimal, they are known to yield near-optimal auctions. And even when they are only approximately optimal, seminal work has identified them as important objects of study owing to their simplicity. In modern discussion of auctions, they are also desirable due to fairness considerations.

While applying existing feed-forward architectures as RegretNet to symmetric auctions is possible, we show in~Section~\ref{sec:NN_arch} that RegretNet struggles to find symmetric auctions, \emph{even when the optimum is symmetric}. To be clear, the architecture's performance is indeed quite close to optimal, but the resulting auction is not ``close to symmetric''. This paper proposes an architecture that outputs a symmetric auction symmetry by design.

 \subsection*{Contributions}
 \label{sec:contributions}

This paper identifies three drawbacks from using the RegretNet architecture when learning with symmetric auctions. First, RegretNet is incapable of finding symmetric auctions when the optimal mechanism is known to be symmetric. Second, RegretNet is sample inefficient, which is not surprising since the architecture does not incorporate any inductive bias. Third, RegretNet is incapable of generalizing to settings with a different number of bidders of objects. In fact, by construction, the solution found by RegretNet can only be evaluated on settings with exactly the same number of bidders and objects of the setting it was trained on.

We address these limitations by proposing a new architecture EquivariantNet, that outputs symmetric auctions. EquivariantNet is an adaption of the deep sets architecture \citep{hartford2018deep} to symmetric auctions.  This architecture is parameter-efficient and is able to  recover some of the optimal results in the symmetric auctions literature.  Our approach outlines three important benefits: 
\begin{itemize}
    \item[--] \textit{Symmetry}: our architecture outputs a symmetric auction by design. It is immune to permutation-sensitivity as defined in Section~\ref{sec:ffpe} which is related to fairness.
    \item[--] \textit{Sample generalization}: Because we use domain knowledge, 
    our architecture converges to the optimum with fewer valuation samples.
    \item[--] \textit{Out-of-setting generalization}: Our architecture does not require hard-coding the number of bidders or items during training --- training our architecture on instances with $n$ bidders and $m$ items produces a well-defined auction even for instances with $n'$ bidders and $m'$ items. Somewhat surprisingly, we show in~\ref{sec:num_exp} some examples where our architecture trained on $1$ bidder with $5$ items generalizes well even to $1$ bidder and $m$ items, for any $m \in \{2,10\}$.
\end{itemize}
 We highlight that the novelty of this paper is not to show that a new architecture is a viable alternative to RegretNet.  Instead we are solving three fundamental limitations we identified for the RegretNet architecture. These three problems are not easy to solve in principle, it is surprising that a change of architecture solves all of them in the context of symmetric auctions. We would also like to emphasize that both RegretNet and EquivariantNet are capable of learning auction with near optimal revenue and negligible regret. It is not possible to significantly outperform RegretNet on these aspects. The way we improve over RegretNet is by having better sample efficiency, out-of-setting generalization and by ensuring that our solutions are exactly equivariant. 


The paper decomposes as follows. Section~\ref{sec:setting} introduces the standard notions of auction design. Section~\ref{sec:NN_arch} presents our permutation-equivariant architecture to encode symmetric auctions. Finally, Section~\ref{sec:num_exp} presents numerical evidence for the effectiveness of our approach.

\subsection*{Related work} 

\paragraph{Auction design and machine learning.} Machine learning and computational learning theory have been used in several ways to design auctions from samples of bidder valuations. Some works have focused sample complexity results for designing optimal revenue-maximizing auctions. This has been established in single-parameter settings \citep{DhangwatnotaiRY15, cole2014sample,morgenstern2015pseudo, medina2014learning,huang2018making, DevanurHP16, HartlineT19,RoughgardenS16, GonczarowskiN17, GuoHZ19},  multi-item auctions \citep{dughmi2014sampling, GonczarowskiW18}, combinatorial auctions \citep{balcan2016sample,morgenstern2016learning,syrgkanis2017sample} and allocation mechanisms \citep{narasimhan2016general}. Machine learning has also been used to optimize different aspects of mechanisms \citep{lahaie2011kernel,dutting2015payment}. All these aforementioned differ from ours as we resort to deep learning for finding optimal auctions. 

\paragraph{Auction design and deep learning.} While \cite{dutting2017optimal} is the first paper to design auctions through deep learning, several other paper followed-up this work. \cite{feng2018deep} extended it to budget constrained bidders, \cite{golowich2018deep} to the facility location problem. \cite{tacchetti2019neural} built architectures based on the Vickrey-Clarke-Groves auctions. Recently, \cite{shen2019automated} and \cite{dutting2017optimal} proposed architectures that \textit{exactly} satisfy incentive compatibility but are specific to \textit{single-bidder} settings. In this paper, we aim at \textit{multi-bidder} settings and build permutation-equivariant networks that return nearly incentive compatibility symmetric auctions. 

\section{Symmetries and learning problem in auction design}\label{sec:setting}

We review the framework of auction design and the problem of finding truthful mechanisms. We then present symmetric auctions and similarly to \cite{dutting2017optimal}, frame auction design as a learning problem.

\subsection{Auction design and symmetries}

\paragraph{Auction design.} We consider the setting of additive auctions with $n$ bidders with $N=\{1,\dots,n\}$ and $m$ items with $M=\{1,\dots,m\}.$ Each bidder $i$ is has value $v_{ij}$ for item $j$, and values the set $S$ of items at $\sum_{j \in S} v_{ij}$. Such valuations are called \emph{additive}, and are perhaps the most well-studied valuations in multi-item auction design~\citep{HartN12,HartN13, LiY13, BabaioffILW14, DaskalakisDT14, HartR15, CaiDW16,daskalakis2017strong, BeyhaghiW19}.

The designer does not know the full valuation profile $V = (v_{ij})_{i\in N, j\in M}$, but just a distribution from which they are drawn. Specifically, the valuation vector of bidder $i$ for each of the $m$ items $\vec{v}_i=(v_{i1}, \dots, v_{im})$ is drawn from a distribution $D_i$ over $\mathbb{R}^m$  (and then, $V$ is drawn from $D:= \times_i D_i$). The designer asks the bidders to report their valuations (potentially untruthfully), then decides on an allocation of items to the bidders and charges a payment to them. 

 
\begin{definition}
 An auction is a pair $(g,p)$ consisting of a randomized allocation rule $g=(g_1,\dots,g_n)$ where $g_i\colon \mathbb{R}^{n\times m}\rightarrow [0,1]^{m}$ such that for all $V$, and all $j$, $\sum_i (g_i(V))_j\leq 1$ and payment rules $p=(p_1,\dots,p_n)$ where $p_i\colon \mathbb{R}^{n\times m}\rightarrow \mathbb{R}_{\geq 0}$ . 
 \end{definition}

 Given reported bids $B=(b_{ij})_{i\in N, j\in M}$, the auction computes an allocation probability $g(B)$ and payments $p(B)$. 
 $[g_i(B)]_j$
 is the probability that bidder $i$ received object $j$ and $p_i(B)$ is the price bidder $i$ has to pay to the mechanism.
 In what follows, $\mathcal{M}$ denotes the class of all possible auctions.
  
 \begin{definition}
 The utility of bidder $i$ is defined by $u_i(\vec{v}_i,B)= \sum_{j=1}^m [g_i(B)]_j v_{ij} -p_i(B).$ 
\end{definition}

Bidders seek to maximize their utility and may report bids that are different from their valuations. Let $V_{-i}$ be the valuation profile without element $\vec{v}_i$, similarly for $B_{-i}$ and $D_{-i}=\times_{j\neq i}D_j$. We aim at auctions that invite bidders to bid their true valuations through the notion of incentive compatibility.

\begin{definition}\label{def:DSIC}
An auction $(g,p)$ is \textit{dominant strategy incentive compatible} (DSIC) if each bidder's utility is maximized by reporting truthfully no matter what the other bidders report. For every bidder $i,$ valuation $\vec{v}_i \in D_i$, bid $\vec{b}_i\hspace{.02cm}'\in D_i$ and bids $B_{-i}\in D_{-i}$, \; $u_i(\vec{v}_i,(\vec{v}_i,B_{-i}))\geq u_i(\vec{v}_i,(\vec{b}_i\hspace{.02cm}',B_{-i})).$ 
\end{definition}
 
Additionally, we aim at auctions where each bidder receives a non-negative utility.
\begin{definition}\label{def:IR}
An auction is \textit{individually rational} (IR) if for all $i\in N, \; \vec{v}_i\in D_i$ and $B_{-i}\in D_{-i},$
\begin{equation}\label{eq:IR_eq}\tag{IR}
   u_i(\vec{v}_i,(\vec{v}_i,B_{-i}))\geq 0.  
\end{equation}
\end{definition}

In a DSIC auction, the bidders have the incentive to truthfully report their valuations and therefore, the revenue on valuation profile $V$ is defined as $\sum_{i=1}^n p_i(V).$ Optimal auction design aims at finding a DSIC auction that maximizes the expected revenue $rev:=\mathbb{E}_{V\sim D}[\sum_{i=1}^n p_i(V)]$. 
  
\paragraph{Linear program.} We frame the problem of optimal auction design as an optimization problem where we seek an auction that minimizes the negated expected revenue among all IR and DSIC auctions. Since there is no known characterization of DSIC mechanisms in the multi-bidder setting, we resort to the relaxed notion of \textit{ex-post regret}. It measures the extent to which an auction violates DSIC, for each bidder.

\begin{definition}
The ex-post regret for a bidder $i$ is the maximum increase in his utility when  considering all his possible bids and fixing the bids of others. For a valuation profile $V$, the ex-post regret for a bidder $i$ is $rgt_{i}(V)=\max_{\vec{v}_i\hspace{.02cm}'\in \mathbb{R}^m} u_i(\vec{v}_i;(\vec{v}_i\hspace{.02cm}',V_{-i}))-u_i(\vec{v}_i;(\vec{v}_i,V_{-i})).$ In particular, DSIC is equivalent to 
\begin{equation}\label{eq:regretDSIC}\tag{IC} 
    rgt_{i}(V)=0, \;  \forall i \in N.
\end{equation}
\end{definition}

Therefore, by setting~\eqref{eq:regretDSIC} and \eqref{eq:IR_eq} as constraints, finding an optimal auction is equivalent to the following linear program
\begin{equation*}\tag{LP}\label{eq:exact_prob}
\begin{aligned}
\hspace{-.41cm}\underset{(g,p)\in \mathcal{M} }{\text{min}}
  - \mathbb{E}_{V\sim D}\left[\sum_{i=1}^n p_i(V)\right] \quad \text{s.t.} \quad 
&   rgt_{i}(V)=0, \hspace{1.55cm} \forall i \in N,\;  \forall V \in D, \\
& u_i(\vec{v}_i,(\vec{v}_i,B_{-i}))\geq 0, \quad  \forall i\in N,\; \vec{v}_i\in D_i, B_{-i}\in D_{-i}.
\end{aligned}
\end{equation*}

\paragraph{Symmetric auctions. } \eqref{eq:exact_prob} is intractable due to the exponential number of constraints. However, in the setting of \textit{symmetric} auctions, it is possible to reduce the search space of the problem as shown in \autoref{prop:equiv_sol}. We first define  the notions of bidder- and item-symmetries. 
\begin{definition}
The valuation distribution $D$ is bidder-symmetric if for any permutation of the bidders ${\varphi_b\colon N\rightarrow N},$ the permuted distribution ${D_{\varphi_b}:=D_{\varphi_b(1)}\times \dots \times D_{\varphi_b(n)}}$ satifies: $D_{\varphi_b}=D$.
\end{definition}
Bidder-symmetry intuitively means that the bidders are a priori indistinguishable (although individual bidders will be different). This holds for instance in auctions where the identity of the bidders is anonymous, or if $D_i = D_j$ for all $i,j$ (bidders are i.i.d.).

\begin{definition}
Bidder $i$'s valuation distribution $D_i$ is item-symmetric if for any items $x_1,\dots,x_m$ and any permutation $\varphi_o\colon M\rightarrow M,$ $D_i(x_{\varphi_o(1)},\dots,x_{\varphi_o(m)} )=D_i(x_1,\dots,x_m )$. 
\end{definition}

Intuitively, item-symmetry means that the items are also indistinguishable but not identical. It holds when the distributions over the items are i.i.d. but this is not a necessary condition. Indeed, the distribution $\{(a,b,c)\in \mathcal{U}(0,1)^{\otimes 3}: a+b+c=1\}$ is not i.i.d. but is item-symmetric.

\begin{definition}\label{ass:bid_it_sym}
An auction is symmetric if its valuation distributions are bidder- and item-symmetric.
\end{definition}

We now define the notion of permutation-equivariance that is important in symmetric auctions.

\begin{definition}
The functions $g$ and $p$ are permutation-equivariant if for any two permutation matrices $\Pi_{n}\in \{0,1\}^{n\times n}$ and $\Pi_{m}\in \{0,1\}^{m\times m}$, and any valuation matrix $V$,  we have $ g(\Pi_{n}\,V\,\Pi_{m})=\Pi_{n}\, g(V)\, \Pi_{m}$ and $p(\Pi_{n}\,V\,\Pi_{m})=\Pi_{n} \, p(V)$.
\end{definition}

\begin{theorem}\label{prop:equiv_sol}
When the auction is symmetric, there exists an optimal solution to \eqref{eq:exact_prob} that is permutation-equivariant.
\end{theorem}

\autoref{prop:equiv_sol} is originally proved in \cite{daskalakis2012symmetries} and its proof is reminded in \autoref{app:thm} for completeness. It encourages to reduce the search space in \eqref{eq:exact_prob} by only optimizing over permutation-equivariant allocations and payments. We implement this idea in Section~\ref{sec:NN_arch} where we build equivariant neural network architectures. Before, we frame auction design as a learning problem. 

\subsection{Auction design as a learning problem}\label{par:auction}
Similarly to \cite{dutting2017optimal}, we formulate auction design as a learning problem. We learn a parametric set of auctions $(g^w,p^w)$ where $w\in \mathbb{R}^d$  parameters and $d\in\mathbb{N}$. Directly solving \eqref{eq:exact_prob} is challenging in practice. Indeed, the auctioneer must have access to the bidder valuations which are unavailable to her. Since she has access to the valuation distribution, we relax~\eqref{eq:exact_prob} and replace the IC constraint for all $V\in D$ by the expected constraint $\mathbb{E}_{V\sim D}[rgt_i(V)]=0$ for all $i\in N.$. In practice, the expectation terms are computed by sampling $L$ bidder valuation profiles drawn i.i.d. from $D$. The empirical ex-post regret for bidder $i$ is
\begin{align}
      \widehat{rgt}_i(w)  = \frac{1}{L}\sum_{\ell=1}^L \max_{\vec{v}_i\hspace{.02cm}'\in \mathbb{R}^m} u_i^w(\vec{v}_i^{(\ell)};(\vec{v}_i\hspace{.02cm}',V_{-i}^{(\ell)}))-u_i(\vec{v}_i^{(\ell)};(\vec{v}_i^{(\ell)},V_{-i}^{(\ell)})), \label{eq:emp_reg}\tag{$\hat{R}$}
\end{align}
 where $u_i^w(\vec{v}_i,B):=\sum_{j=1}^m [g_i^w(B)]_j v_{ij} -p_i^w(B)$ is the utility of bidder $i$ under the parametric set of auctions  $(g^w,p^w)$. Therefore, the learning formulation of \eqref{eq:exact_prob} is
 \vspace{-0.5em}
 \begin{equation}\label{eq:empirical_prob}\tag{$\widehat{\mathrm{LP}}$}
\begin{aligned}
\underset{w\in \mathbb{R}^d}{\text{min}}
-\frac{1}{L}\sum_{\ell=1}^L \sum_{i=1}^n p_i^w(V^{(\ell)}) \quad \text{s.t.}\quad \widehat{rgt}_i(w)=0,\; \forall i \in N. 
\end{aligned}
\end{equation}

\cite{dutting2017optimal} justify the validity of this reduction from \eqref{eq:exact_prob} to \eqref{eq:empirical_prob} by showing that the gap between the expected regret and the empirical regret is small as the number of samples increases. Additionally to being DSIC, the auction must satisfy IR. The learning problem \eqref{eq:empirical_prob} does not ensure this but we will show how to include this requirement in the architecture in \autoref{sec:NN_arch}.

\section{Permutation-equivariant neural network architecture}\label{sec:NN_arch}

We first show that feed-forward architectures as RegretNet \citep{dutting2017optimal} may struggle to find a symmetric solution in auctions where the optimal solution is known to be symmetric.
We then describe our neural network architecture, EquivariantNet that learns symmetric auctions. EquivariantNet is build using exchangeable matrix layers \citep{hartford2018deep}. 

\subsection{Feed-forward nets and permutation-equivariance}\label{sec:ffpe}

In the following experiments we use the RegretNet architecture  with the exact same training procedure and parameters as found in \cite{dutting2017optimal} .

\paragraph{Permutation-sensitivity.} Given $L$ bidders valuation samples $\{B^{(1)},\dots,B^{(L)}\} \in \mathbb{R}^{n\times m}$,  we generate for each bid matrix $B^{(\ell)} $ all its possible permutations $B_{\Pi_n,\Pi_m}^{(\ell)}:=\Pi_{n}B^{(\ell)}\Pi_{m},$ where ${\Pi_{n}\in \{0,1\}^{n\times n}}$ and ${\Pi_{m}\in \{0,1\}^{m\times m}}$ are permutation matrices.  We then compute the revenue for each one of these bid matrices and obtain a revenue matrix ${R\in \mathbb{R}^{n!m!\times L}}$. Finally, we compute $h_R \in \mathbb{R}^{L}$ where $[h_R]_j = \max_{i\in [n!m!]} R_{ij}-\min_{i\in [n!m!]}R_{ij}$. The distribution given by the entries of $h_R$ is a measure of how close the auction is to permutation-equivariance.
A symmetric mechanism satisfies $h_R = (0,\dots,0)^{\top}$. 
Our numerical investigation considers the following auction settings:
 \begin{itemize}
     \item[--] (I) One bidder and two items, the item values are drawn from $\mathcal{U}[0,1].$ Optimal revenue: 0.55 \cite{manelli2006bundling}.
    \item[--](II) Four bidders and five items, the item values are drawn from $\mathcal{U}[0,1].$ 
 \end{itemize}

\begin{figure}[h]
\begin{minipage}{.5\textwidth}
 \centering
  \includegraphics[width=1.05\linewidth]{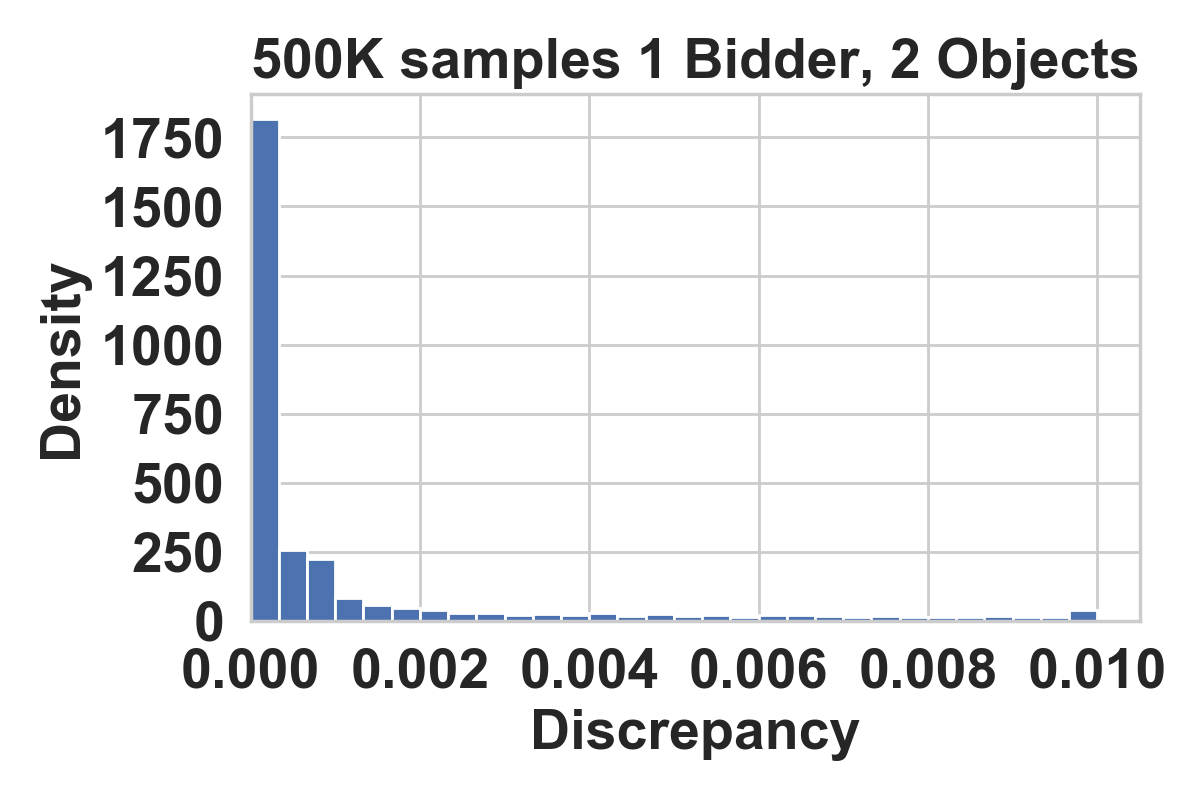}
\vspace*{-.3cm}
\captionsetup{labelformat=empty}
  \caption{(a)}
\end{minipage}%
\begin{minipage}{.5\textwidth}
 \centering
  \includegraphics[width=1.05\linewidth]{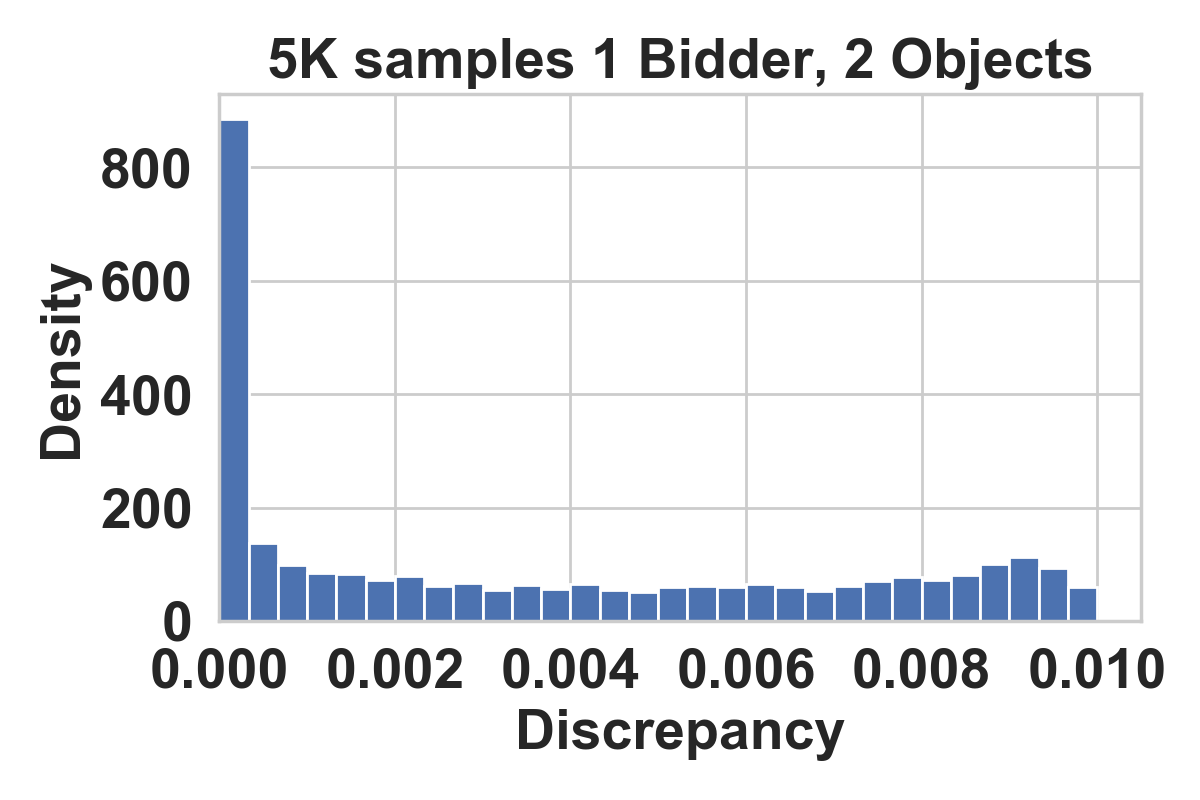}
  \vspace*{-.3cm}
  \captionsetup{labelformat=empty}
\caption{(b)}
\end{minipage}
\begin{minipage}{.5\textwidth}
 \centering
  \includegraphics[width=1.05\linewidth]{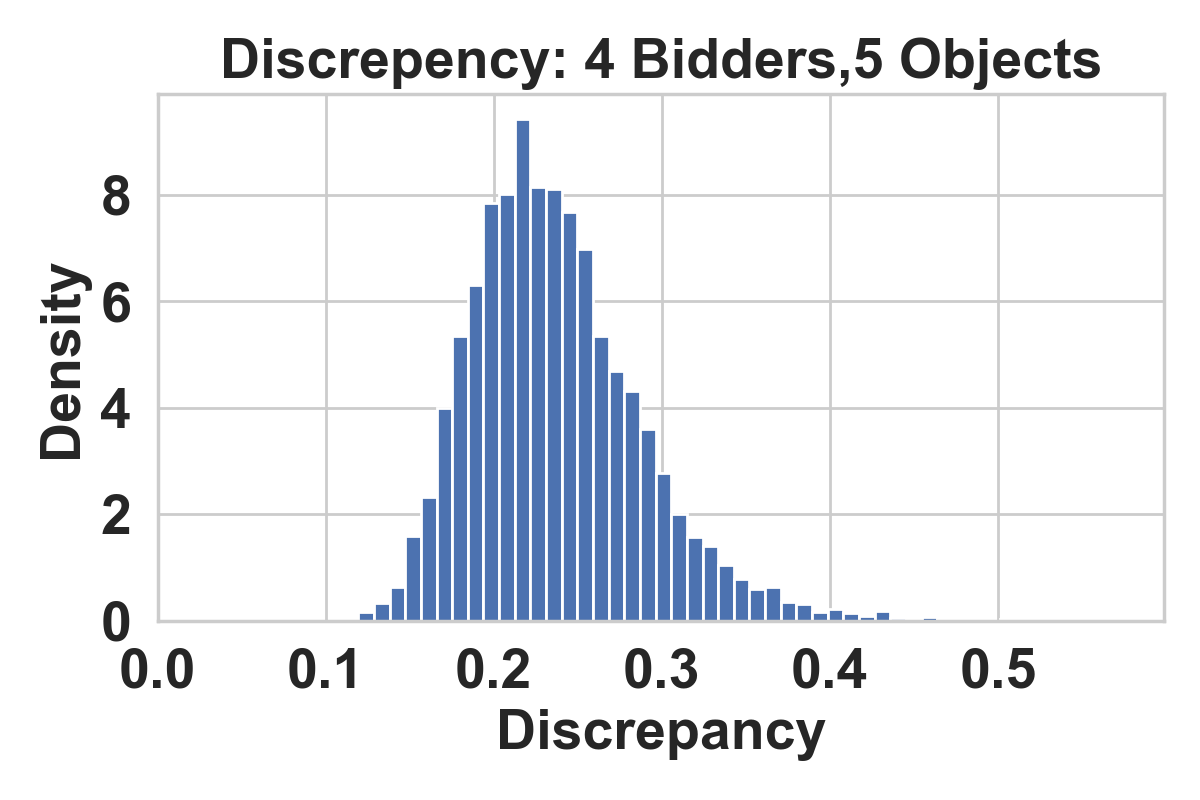}
\vspace*{-.6cm}
\captionsetup{labelformat=empty}
  \caption{(c)}
\end{minipage}%
\begin{minipage}{.5\textwidth}
 \centering
  \includegraphics[width=1.05\linewidth]{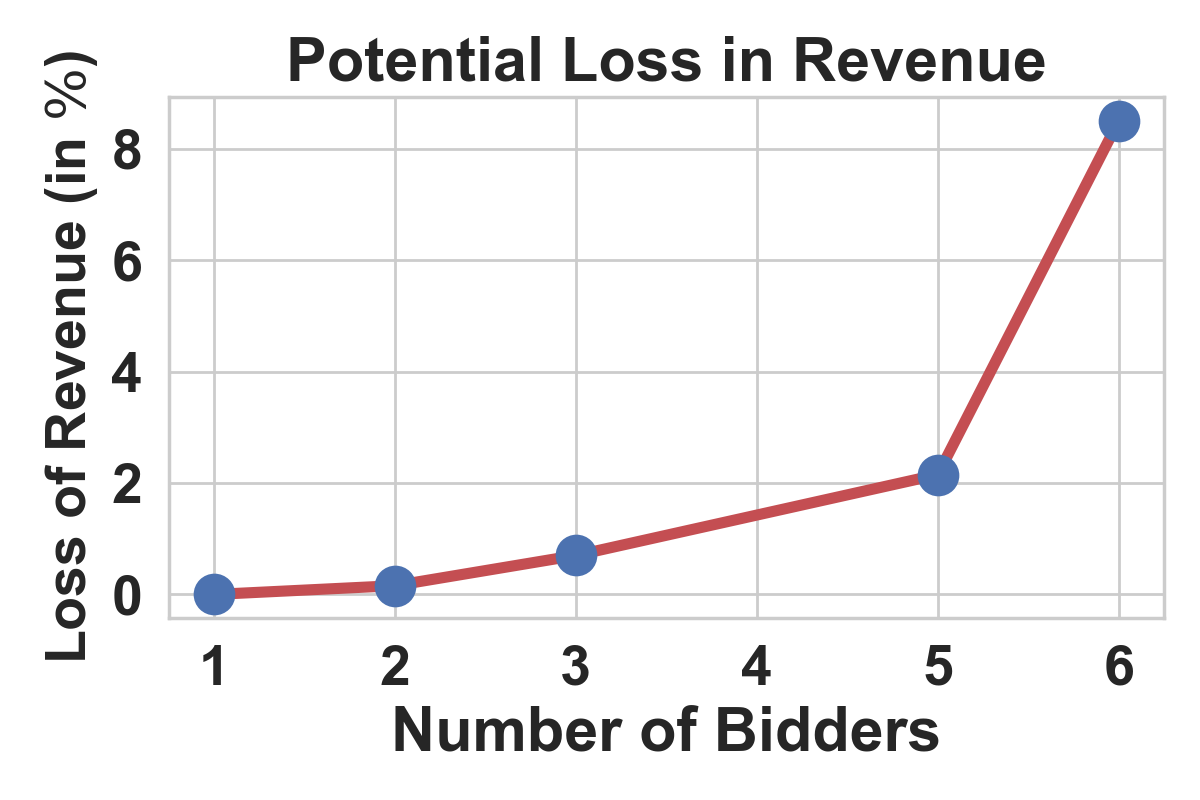}
  \vspace*{-.6cm}
  \captionsetup{labelformat=empty}
\caption{(d)}
\end{minipage}
\setcounter{figure}{0}    
\caption{(a)-(b): Distribution $h_R$ when varying the number of training samples (a) 500 000 (b) 5000 samples. (c): Histogram of the distribution $h_R$ for setting (II). (d): Maximum revenue loss when varying the number of bidders for setting (III$_n$)}.\label{fig:PE_FF}
\end{figure}

\autoref{fig:PE_FF}~(a)-(b) presents the distribution of $h_R$ of the optimal auction learned for setting (I) when varying the number of samples $L.$ When $L$ is large, the distribution is almost concentrated at zero and therefore the network is almost able to recover the permutation-equivariant solution. When $L$ is small, $h_R$ is less concentrated around zero and therefore, the solution obtained is non permutation-equivariant. \\

As the problem's dimensions increase, this lack of permutation-invariance becomes more dramatic. \autoref{fig:PE_FF}~(c) shows $h_R$ for the optimal auction mechanism learned for setting (II) when trained with $5\cdot 10^5$ samples. Contrary to (I), almost no entry of $h_R$ is located around zero, they are concentrated around between $0.1$ and $0.4$ i.e.\ between 3.8\% and 15\% of the estimated optimal revenue. 

\paragraph{Exploitability.} Finally, to highlight how important equivariant solutions are, we analyze the worst-revenue loss that the auctioneer can incur when the bidders act adversarially. Indeed, since different permutations can result in different revenues for the auction, cooperative bidders could pick among the $n!$ possible permutations of their labels the one that minimized the revenue of the mechanism and present themselves in that order. Instead of getting a revenue of $R_{opt} = \mathbb{E}_{V\sim D}\left[\sum_{i=1}^n p_i(V)\right]$, the auctioneer would get a revenue of $R_{adv} = \mathbb{E}_{V\sim D}\left[\min_{\Pi_n}\{\sum_{i=1}^n p_i(\Pi_n V)\}\right]$. The percentage of revenue loss is given by $ l = 100 \times \frac{R_{opt}-R_{adv}}{R_{opt}}$. We compute $l$ in 
in the following family of settings: 
\begin{itemize}
    \item[--] (III$_n$) $n$ additive bidders and ten item where the item values are drawn from $\mathcal{U}[0,1].$
\end{itemize}
In \autoref{fig:PE_FF}~(d) we plot $l(n)$ the loss in revenue as a function of $n$. As the number of bidders increases, the loss becomes more substantial getting over the 8\% with only 6 bidders. \\

While it is unlikely that all the bidders will collide and exploit the bidding mechanism in real life, these investigations of permutation sensitivity and exploitability give us a sense of how far the solutions found by RegretNet are from being bidder-symmetric. The underlying real problem with non bidder-symmetric solution has to do with fairness. RegretNet finds mechanisms that do not treat all bidders equally. Their row number in the bid matrix matters, two bidders with the same bids will not get the same treatment.  If the mechanism is equivariant however, all bidders will be treated equally by design, there are no biases or special treatments.  Aiming for symmetric auctions is important and
to this end, we design a permutation-equivariant architecture.

\begin{figure*}[t]
  \centering
  \includegraphics[width=\linewidth]{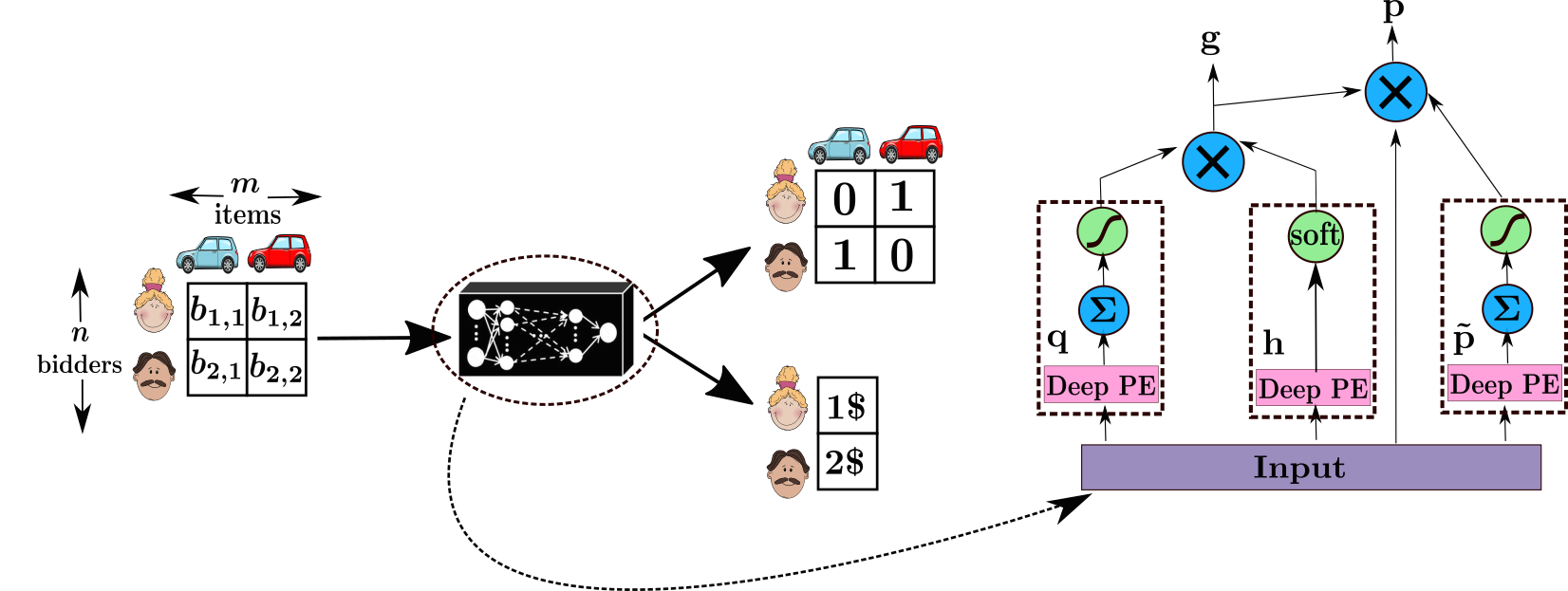}
  \caption{\textit{Left:} Auction design setting. \textit{Right:} EquivariantNet: Deep permutation-equivariant architecture for auction design. Deep PE denotes the deep permutation-equivariant architecture described in \autoref{sec:PE_net}, $\sum$ the sum over rows/columns operations, $\times$ the multiplication operations, soft stands for soft-max and the curve for sigmoid. The network outputs an allocation $g$ and a payment~$p.$
  }\label{fig:arch}
\end{figure*} 

\subsection{Architecture for symmetric auctions (EquivariantNet)}\label{sec:PE_net}
 Our input is a bid matrix $B = (b_{i,j})\in \mathbb{R}^{n\times m}$ drawn from a bidder-symmetric and item-symmetric distribution. We aim at learning a randomized allocation neural network $g^w\colon \mathbb{R}^{n\times m}\rightarrow [0,1]^{n\times m}$ and a payment network $p^w \colon \mathbb{R}^{n\times m}\rightarrow \mathbb{R}_{\geq 0}^n$. The symmetries of the distribution from which $B$ is drawn and \autoref{prop:equiv_sol} motivates us to model $g^w$ and $p^w$ as permutation-equivariant functions. To this end, we use \textit{exchangeable matrix layers} \citep{hartford2018deep} and their definition is reminded in \autoref{app:permeq_net}. We now describe the three modules of the allocation and payment networks \autoref{fig:arch}.\\
 
The first network outputs a vector $q^w(B) \in [0,1]^m$ such that entry $q_j^w(B)$ is the probability that item $j$ is allocated to any of the $n$ bidders. The architecture consists of three modules. The first one is a deep permutation-equivariant network with tanh activation functions. The output of that module is a matrix $Q\in \mathbb{R}^{n\times m}$. The second module transforms $Q$ into a vector $\mathbb{R}^{m}$ by taking the average over the rows of $Q$. We finally apply the sigmoid function to the result to ensure that $q^w(B) \in [0,1]^m$. This architecture ensures that $q^w(B)$ is invariant with respect to bidder permutations and equivariant with respect to items permutations.\\

The second network outputs a matrix $h(B) \in [0,1]^{n\times m}$ where $h_{ij}^w$ is the probability that item $j$ is allocated to bidder $i$ conditioned on item $j$ being allocated. The architecture consists of a deep permutation-equivariant network with tanh activation functions followed by softmax activation function so that $\sum_{i=1}^n h_{ij}^w(B) =  1$. This architecture ensures that $q^w$ equivariant with respect to object and bidder permutations.\\

By combining the outputs of $q^w$ and $h^w,$ we compute the allocation function $g^w\colon \mathbb{R}^{n\times m}\rightarrow [0,1]^{n\times m}$ where $g_{ij}(B)$ is the probability that the allocated item $j$ is given to bidder $i$. Indeed, using conditional probabilities, we have $g_{ij}^w(B)=q_j^w(B) h_{ij}^w(B).$ Note that $g^w$ is a permutation-equivariant function.\\

The third network outputs a vector $p(B) \in \mathbb{R}_{\geq 0}^n$ where $\tilde{p}_i^w $ is the fraction of bidder's $i$ utility that she has to pay to the mechanism. Given the allocation function $g^w$, bidder $i$ has to pay an amount  $p_i = \tilde{p}_i(B)\sum_{j=1}^m g_{ij}^w(B)B_{ij}$. 
Individual rationality is ensured by having $\tilde{p}_i\in [0,1]$.
The architecture of $\tilde{p}^w$ is almost similar to the one of $q^w$. Instead of averaging over the rows of the matrix output by the permutation-equivariant architecture, we average over the columns.

\subsection{Optimization and training}

The optimization and training procedure of EquivariantNet is similar to \cite{dutting2017optimal}. For this reason, we briefly mention the outline of this procedure and remind the details in \autoref{app:opt_train}. We apply the augmented Lagrangian method to  \eqref{eq:emp_reg}. The Lagrangian with a quadratic penalty is:

\begin{align*}
    \mathcal{L}_{\rho}(w;\lambda)&=-\frac{1}{L}\sum_{\ell=1}^L \sum_{i\in N} p_i^w(V^{(\ell)})+\sum_{i\in N}\lambda_i\widehat{rgt}_i(w)+\frac{\rho}{2}\sum_{i \in N}\left(\widehat{rgt}_i(w)\right)^2, 
\end{align*}
where $\lambda \in\mathbb{R}^n$ is a vector of Lagrange multipliers and $\rho>0$ is a fixed parameter controlling the weight of the quadratic penalty. The solver alternates between the updates on model parameters and Lagrange multipliers:  $w^{new}\in\mathrm{argmax}_w \mathcal{L}_{\rho}(w^{old},\lambda^{old})$ and   $\lambda_i^{new}=\lambda_i^{old}+\rho\cdot \widehat{rgt}_i(w^{new}), $ $\forall i \in N.$

\section{Experimental Results}\label{sec:num_exp}

We start by showing the effectiveness of our architecture in symmetric and asymmetric auctions. We then highlight its sample-efficiency for training and its ability to extrapolate to other settings. More details about the setup and training can be found in \autoref{app:opt_train} and \autoref{app:setup}.

\paragraph{Evaluation.} In addition to the revenue of the learned auction on a test set, we also evaluate the corresponding empirical average regret over bidders $\widehat{rgt} = \frac{1}{n} \sum_{i=1}^n \widehat{rgt}_i $. We evaluate these terms by running gradient ascent on $v_i'$ with a step-size of $0.001$ for $\{300,500\}$ iterations (we test $\{100,300\}$ different random initial $v_i'$ and report the one achieves the largest regret).

\paragraph{Known optimal solution.} We first consider instances of single bidder multi-item auctions where the optimal mechanism is known to be symmetric. While independent private value auction as (I) fall in this category, the following item-asymmetric auction has surprisingly an optimal symmetric solution.
\begin{itemize}
    \item (IV) One bidder and two items where the item values are independently drawn according to the probability densities $f_1(x)=5/(1+x)^6$ and $f_2(y)=6/(1+y)^7.$ Optimal solution in \cite{daskalakis2017strong}. 
 \end{itemize}

\begin{figure}[ht]
\begin{minipage}{.45\textwidth}
{\hspace{-.3cm}\resizebox{1\columnwidth}{!}{
\begin{tabular}{ cccc } 
\toprule
 Dist. & $rev$ & $rgt$ & OPT  \\
\midrule
  (I) & 0.551 & 0.00013  &0.550      \\ 
 (IV) & 0.173 & 0.00003 & 0.1706    \\
 (V) &  0.873 & 0.001 & 0.860 \\ 
\bottomrule
\end{tabular}}}
\vspace*{.7cm}
\captionsetup{labelformat=empty}
\caption{(a)}
\end{minipage} \hspace{3em}
\begin{minipage}{.45\textwidth}
{\hspace{-.3cm}\resizebox{1.05\columnwidth}{!}{
\begin{tabular}{ ccccc } 
\toprule
\multicolumn{1}{c}{ }
& \multicolumn{2}{c}{ EquivariantNet}& \multicolumn{2}{c}{ RegretNet} \\
\cmidrule(lr){2-3}
\cmidrule(lr){4-5}
$\lambda_2$ & $rev$ & $rgt$ & $rev_F$ & $rgt_F$ \\
\cmidrule(lr){1-1}
\cmidrule(lr){2-2}
\cmidrule(lr){3-3}
\cmidrule(lr){4-4}
\cmidrule(lr){5-5}
 0.01 &  0.37 & 0.0006 & 0.39 & 0.0003\\
 0.1 & 0.41 &0.0004 &0.41 &0.0007 \\
 1 & 0.86 &0.0005 &0.84  & 0.0012 \\
 10 & 3.98 &0.0081 & 3.96 &0.0056\\
\bottomrule
\end{tabular}}}
\captionsetup{labelformat=empty}
\caption{(b)}
\end{minipage}%
\setcounter{figure}{2}    
\caption{(a): Test revenue and regret found by EquivariantNet for settings (I), (IV) and (V). For seeting (V)  OPT is the optimal revenue from VVCA and AMA$_{\mathrm{bsym}}$ families of auctions \citep{sandholm2015automated}. For settings (I) and (IV), OPT is the theoretical optimal revenue. (b): Test revenue/regret for setting (VI) when varying $\lambda_2$ ($\lambda_1=1$). $rev_F$ and $rgt_F$ are computed with RegretNet.}\label{fig:tables}

\end{figure}

\begin{figure}[ht]
\begin{minipage}{.48\textwidth}
\hspace{-.3cm}
  \includegraphics[width=1.0\linewidth]{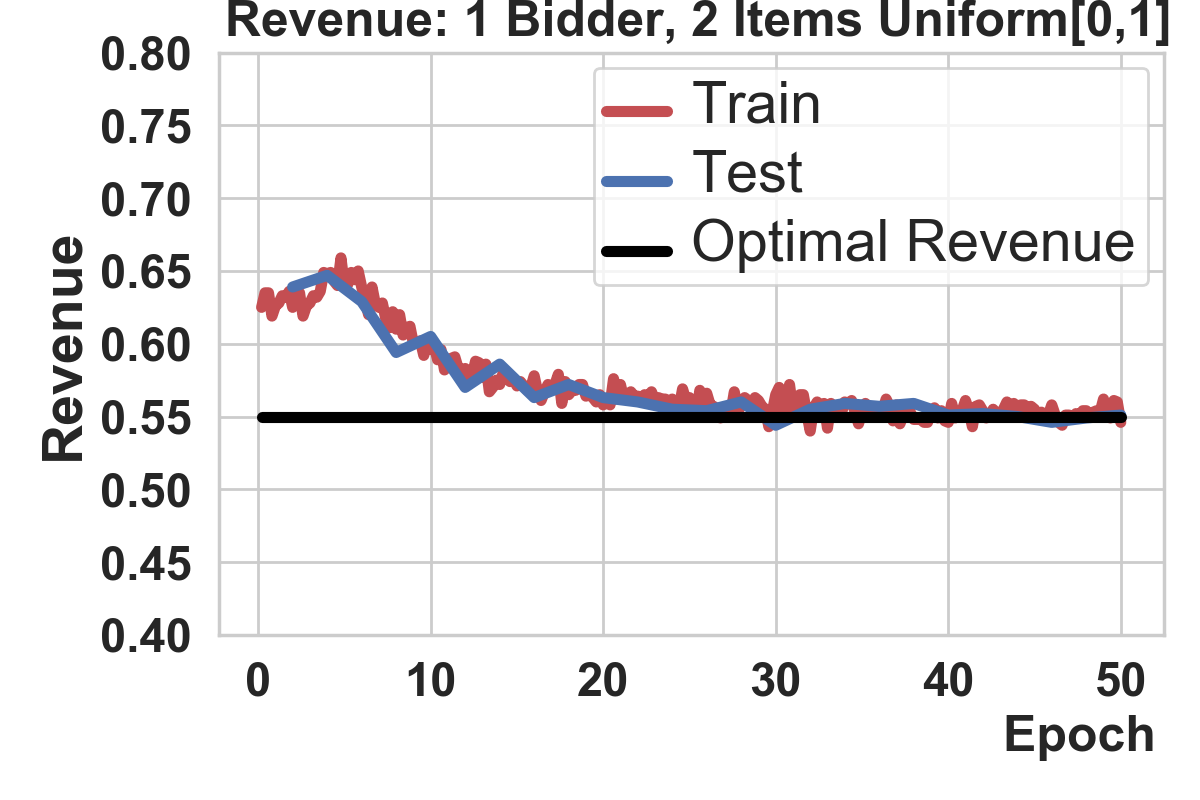}
  \captionsetup{labelformat=empty}
    \vspace*{-.2cm}
  \caption{\hspace{-.2cm}(a)}
\end{minipage}
\begin{minipage}{.48\textwidth}
\vspace{-0.3em}
  \includegraphics[width=1.08\linewidth]{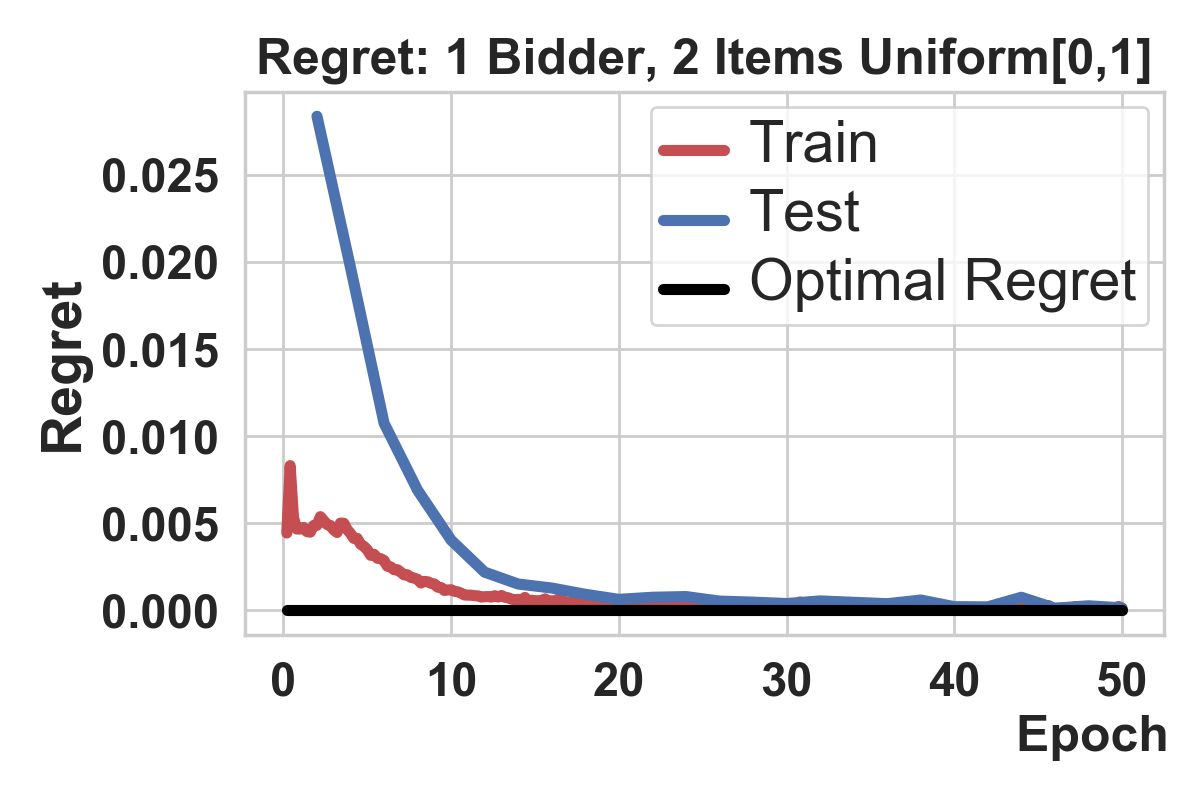}
  \captionsetup{labelformat=empty}
\caption{\hspace{.8cm}(b)}
\end{minipage}
\setcounter{figure}{3}    
\caption{ Train/test revenue (a) and regret (b) as a function of epochs for setting (I) for EquivariantNet. The revenue converges to the theoretical optimum (0.55) and the regret converges to 0. EquivariantNet recovers the optimal mechanism.  }\label{fig:symm_auc}
\end{figure}

The two first lines in \autoref{fig:tables}(a) report the revenue and regret of the mechanism learned by our model. The revenue  is very close to the optimal one, and the regret is  negligible. Remark that the learned auction may achieve a revenue slightly above the optimal incentive compatible auction. This is possible because although small, the regret is non-zero. \autoref{fig:symm_auc}(a)-(b) presents a plot of revenue and regret as a function of training epochs for the setting (I).

\paragraph{Unknown optimal solution.}
Our architecture is also able to recover a permutation-equivariant solution in settings for which the optimum is not known analytically such as:

\begin{itemize}
    \item[--] (V) Two additive bidders and two items where bidders draw their value for each item from $\mathcal{U}[0,1].$  
\end{itemize}

 We compare our solution to the optimal auctions from the VVCA and AMA$_{\mathrm{bsym}}$ families of incentive compatible auctions from \citep{sandholm2015automated}. The last line of \autoref{fig:tables}(a) summarizes our results.
 
 \vspace{-.2cm}
 
\paragraph{Non-symmetric optimal solution.} Our architecture returns satisfactory results in asymmetric auctions. (VI) is a setting where there may not be permutation-equivariant solutions. 

\begin{itemize}
    \item[--] (VI)  Two bidders and two items where the item values are independently drawn according to the probability densities $f_1(x)=\lambda_1^{-1}e^{-\lambda_1 x}$ and $f_2(y)=\lambda_2^{-1}e^{-\lambda_2 y},$ where $\lambda_1, \lambda_2>0.$ 
 \end{itemize}
 
\autoref{fig:tables}(b) shows the revenue and regret of the final auctions learned for setting (VI). When $\lambda_1=\lambda_2,$ the auction is symmetric and so, the revenue of the learned auction is very close to the optimal revenue, with negligibly small regret. However, as we increase the gap between $\lambda_1$ and $\lambda_2,$ the asymmetry becomes dominant and the optimal auction does not satisfy permutation-equivariance. We remark that our architecture does output a solution with near-optimal revenue and small regret.

\paragraph{Sample-efficiency.} Our permutation-equivariant architecture exhibits solid generalization properties when compared to the feed-forward architecture RegretNet. When enough data is available at training, both architectures generalize well to unseen data and the gap between the training and test losses goes to zero. However, when fewer training samples are available, our equivariant architecture generalizes while RegretNet struggles to. This may be explained by the inductive bias in our architecture. We demonstrate this for auction (V) with a training set of $20$ samples and  plot the training and test losses as a function of time (measures in epochs) for both architectures in \autoref{fig:general_exp}(a).


\begin{figure}[h]
\centering
\begin{minipage}{.6\textwidth}
  \includegraphics[width=\linewidth]{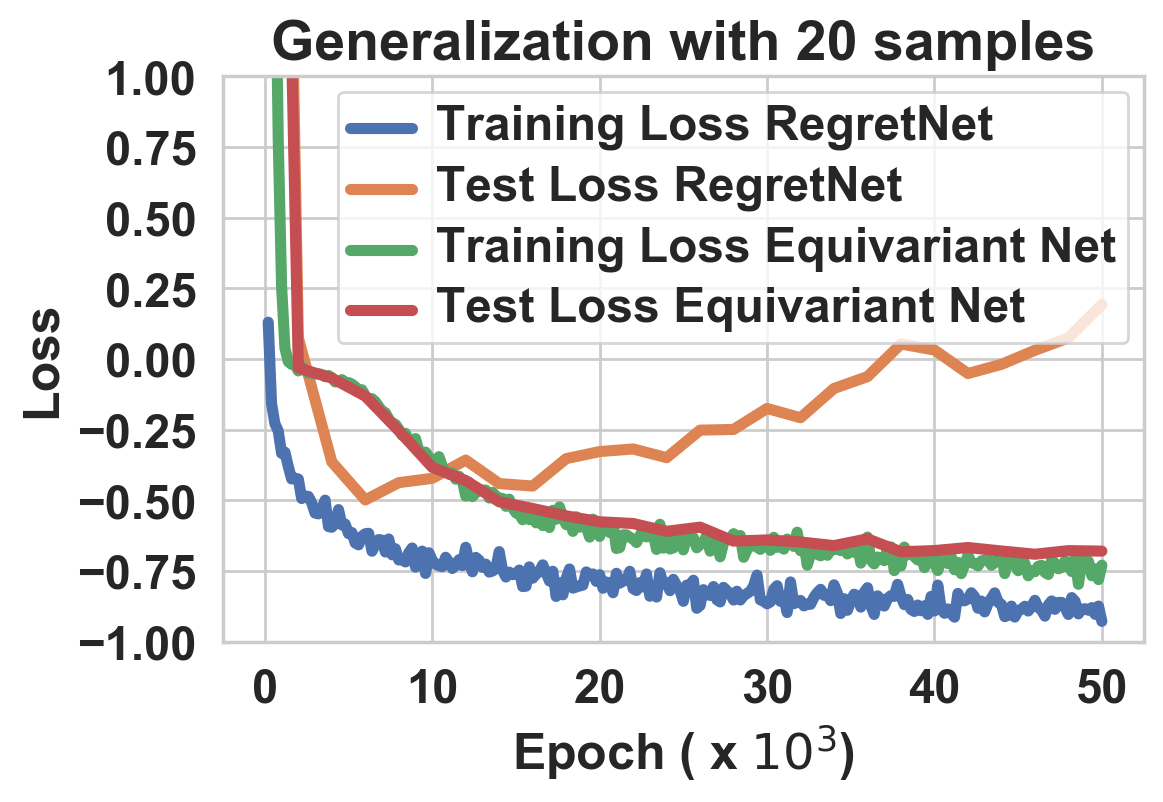}
\vspace*{-.4cm}
\captionsetup{labelformat=empty}
\end{minipage}
\setcounter{figure}{4}    
\caption{Train and test losses (the Lagrangian) for setting (V) with 20 training samples. RegretNet and EquivariantNet both achieve small losses on the training set, only EquivariantNet is generalizes to the testing set.   }\label{fig:general_exp}
\end{figure}

\paragraph{Out-of-setting generalization.} The number of parameters in our permutation-equivariant architecture does not depend on the size of the input. Given an architecture that was trained on samples of size $(n,m)$, it is possible to evaluate it on samples of any size $(n',m')$ (More details in \autoref{app:permeq_net}). This  evaluation is not well defined for feed-forward architectures where the dimension of the weights depends on the input size. We use this advantage to check whether models trained in a fixed setting perform well in totally different ones. 
 
\begin{itemize}
\item[--]($\alpha$) Train an equivariant architecture on 1 bidder, 5 items and test it on 1 bidder, $n$ items for $n = 2 \cdots 10$. All the items values are sampled independently from $\mathcal{U}[0,1].$ 

\item[--]($\beta$) Train an equivariant architecture on 2 bidders, 3 objects and test it on 2 bidders, $n$ objects for $n= 2 \cdots 6$.  All the items values are sampled independently from $\mathcal{U}[0,1].$ 
\end{itemize}
\autoref{fig:general_exp_out}(a)-(b) reports the test revenue that we get for different values of $n$ in ($\alpha)$ and ($\beta$) and compares it to the empirical optimal revenue. Our baseline for that is RegretNet. Surprisingly, our model does generalize well. It is worth mentioning that knowing how to solve a larger problem such as $1 \times 5$ does not automatically result in a capacity to solve a smaller one such as $1\times 2$; the generalization does happen on both ends. Our approach looks promising regarding out of setting generalization. It generalizes well when the number of objects varies and the number of bidders remain constants. However, generalization to settings where the number of bidders varies is more difficult due to the complex interactions between bidders. We do not observe good generalization with our current method. 

\begin{figure}[h]
\begin{minipage}{.5\textwidth}
\centering
  \includegraphics[width=\linewidth]{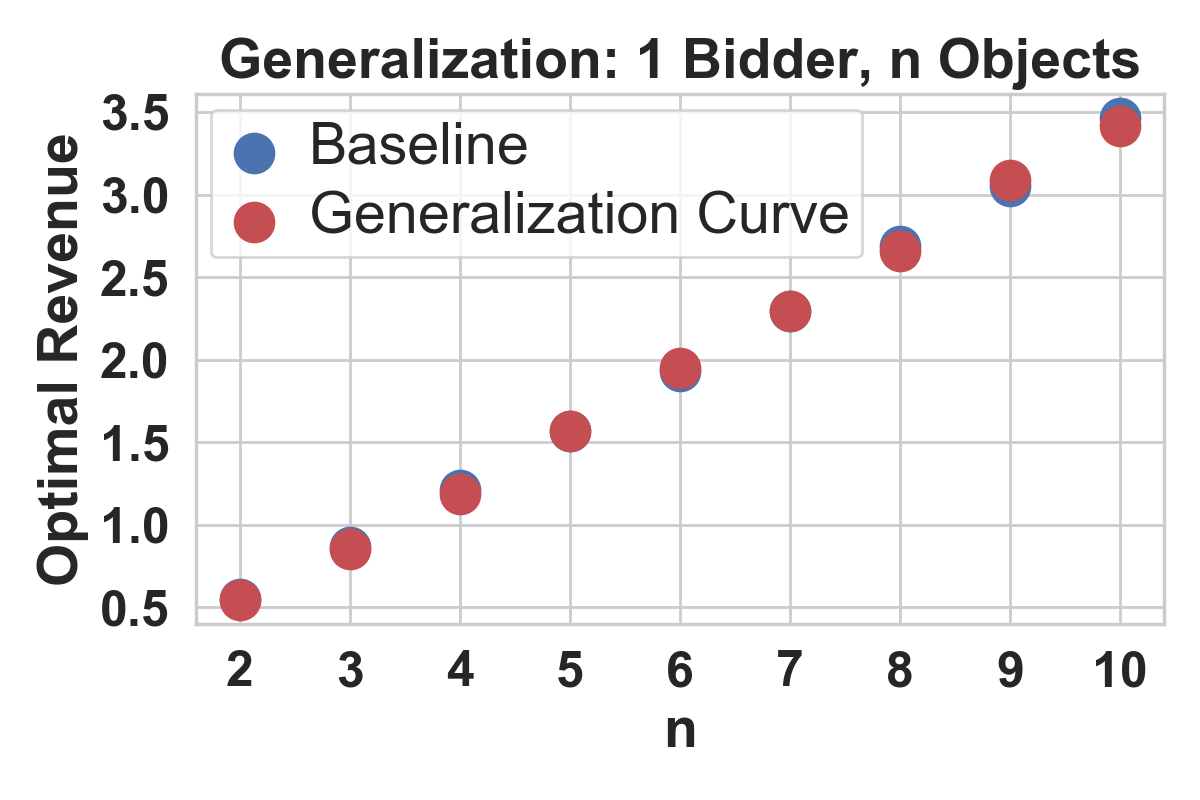}
\captionsetup{labelformat=empty}
\vspace*{-.4cm}
  \caption{(a)}
\end{minipage}%
\begin{minipage}{.5\textwidth}
\centering
  \includegraphics[width=\linewidth]{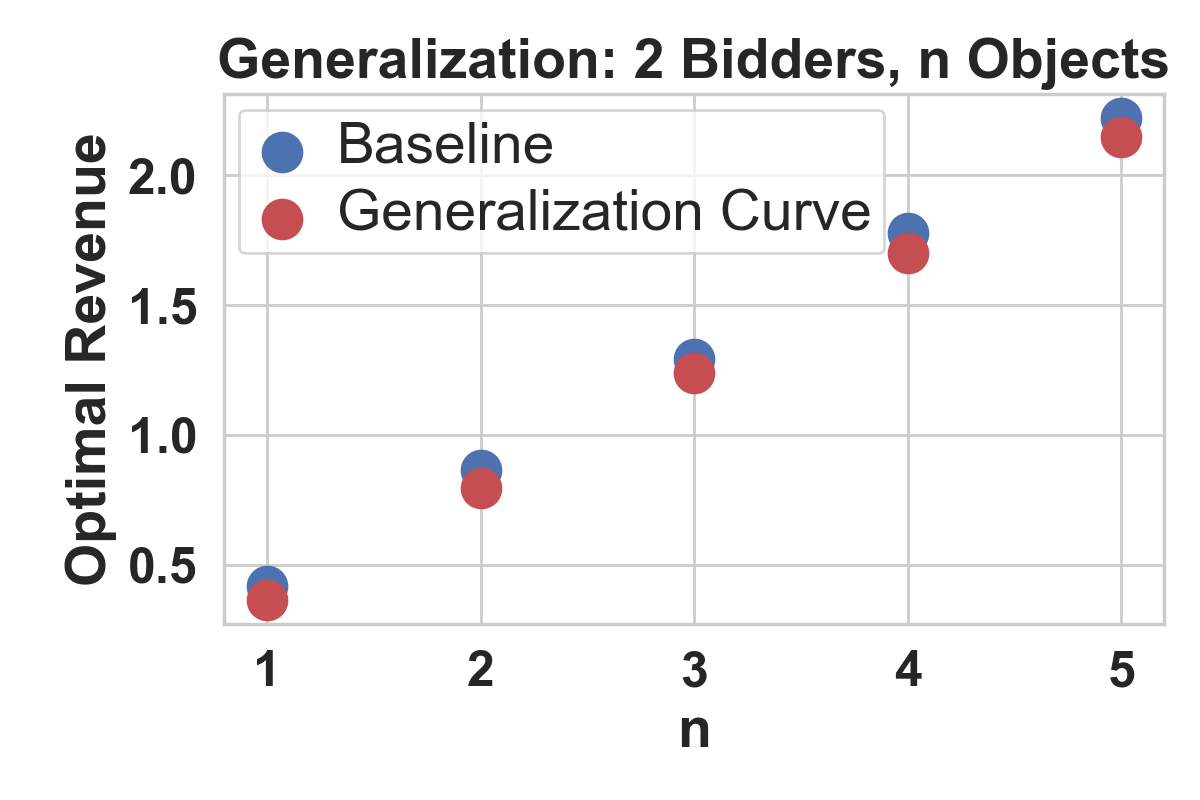}
  \vspace*{-.4cm}
  \captionsetup{labelformat=empty}
\caption{(b)}
\end{minipage}
\setcounter{figure}{5}    
\caption{Figure 5:Generalization revenue of EquivariantNet in experiment $(\alpha)$ and $(\beta).$ Each baseline point is computed using a RegretNet architecture trained from scratch.  }\label{fig:general_exp_out}
\end{figure}

\section*{Conclusion}

We have explored the effect of adding domain knowledge in neural network architectures for auction design. We built a permutation-equivariant architecture to design symmetric auctions and highlighted its multiple advantages. It recovers several known optimal results and provides competitive results in asymmetric auctions. Compared to fully connected architectures, it is more sample efficient and is able to generalize to settings it was not trained on. In a nutshell, this paper insists on the importance of bringing domain-knowledge to the deep learning approaches for auctions.\\
 
Our architecture presents some limitations. It assumes that all the bidders and items are permutation-equivariant. However, in some real-world auctions, the item/bidder-symmetry only holds for a group of bidders/items. More advanced architectures such as Equivariant Graph Networks \cite{maron2018invariant} may solve this issue. Another limitation is that we only consider additive valuations. An interesting direction would be to extend our approach to other settings as unit-demand or combinatorial auctions.

\paragraph{Acknowledgements.}
We would like to thank Ryan P. Adams, Devon Graham and Andrea Tacchetti
for helpful discussions. The work of Jad Rahme was funded by a Princeton SEAS Innovation Grant. Samy Jelassi and Joan Bruna's work are partially supported by the Alfred P. Sloan Foundation, NSF RI-1816753, NSF CAREER CIF 1845360, NSF CHS-1901091, Samsung Electronics, and the Institute for Advanced Study. The work of S. Matthew Weinberg was supported by NSF CCF-1717899.

\pagebreak

\bibliography{biblio_new}

\begin{thebibliography}{56}
\providecommand{\natexlab}[1]{#1}
\providecommand{\url}[1]{\texttt{#1}}
\expandafter\ifx\csname urlstyle\endcsname\relax
  \providecommand{\doi}[1]{doi: #1}\else
  \providecommand{\doi}{doi: \begingroup \urlstyle{rm}\Url}\fi

\bibitem[Alaei(2011)]{Alaei11}
Saeed Alaei.
\newblock {Bayesian Combinatorial Auctions: Expanding Single Buyer Mechanisms
  to Many Buyers}.
\newblock In \emph{the 52nd Annual IEEE Symposium on Foundations of Computer
  Science (FOCS)}, 2011.

\bibitem[Alaei et~al.(2012)Alaei, Fu, Haghpanah, Hartline, and
  Malekian]{AlaeiFHHM12}
Saeed Alaei, Hu~Fu, Nima Haghpanah, Jason Hartline, and Azarakhsh Malekian.
\newblock {Bayesian Optimal Auctions via Multi- to Single-agent Reduction}.
\newblock In \emph{the 13th ACM Conference on Electronic Commerce (EC)}, 2012.

\bibitem[Alaei et~al.(2013)Alaei, Fu, Haghpanah, and Hartline]{AlaeiFHH13}
Saeed Alaei, Hu~Fu, Nima Haghpanah, and Jason Hartline.
\newblock {The Simple Economics of Approximately Optimal Auctions}.
\newblock In \emph{the 54th Annual IEEE Symposium on Foundations of Computer
  Science (FOCS)}, 2013.

\bibitem[Alaei et~al.(2015)Alaei, Hartline, Niazadeh, Pountourakis, and
  Yuan]{AlaeiHNPY15}
Saeed Alaei, Jason~D. Hartline, Rad Niazadeh, Emmanouil Pountourakis, and Yang
  Yuan.
\newblock Optimal auctions vs. anonymous pricing.
\newblock \emph{CoRR}, abs/1507.02615, 2015.
\newblock URL \url{http://arxiv.org/abs/1507.02615}.

\bibitem[Babaioff et~al.(2014)Babaioff, Immorlica, Lucier, and
  Weinberg]{BabaioffILW14}
Moshe Babaioff, Nicole Immorlica, Brendan Lucier, and S.~Matthew Weinberg.
\newblock A simple and approximately optimal mechanism for an additive buyer.
\newblock \emph{SIGecom Exchanges}, 13\penalty0 (2):\penalty0 31--35, 2014.
\newblock \doi{10.1145/2728732.2728736}.
\newblock URL \url{http://doi.acm.org/10.1145/2728732.2728736}.

\bibitem[Balcan et~al.(2016)Balcan, Sandholm, and Vitercik]{balcan2016sample}
Maria-Florina~F Balcan, Tuomas Sandholm, and Ellen Vitercik.
\newblock Sample complexity of automated mechanism design.
\newblock In \emph{Advances in Neural Information Processing Systems}, pages
  2083--2091, 2016.

\bibitem[Beyhaghi and Weinberg(2019)]{BeyhaghiW19}
Hedyeh Beyhaghi and S.~Matthew Weinberg.
\newblock Optimal (and benchmark-optimal) competition complexity for additive
  buyers over independent items.
\newblock In \emph{Proceedings of the 51st ACM Symposium on Theory of Computing
  Conference (STOC)}, 2019.

\bibitem[Cai et~al.(2012{\natexlab{a}})Cai, Daskalakis, and Weinberg]{CaiDW12a}
Yang Cai, Constantinos Daskalakis, and S.~Matthew Weinberg.
\newblock An algorithmic characterization of multi-dimensional mechanisms.
\newblock In \emph{Proceedings of the 44th Symposium on Theory of Computing
  Conference, {STOC} 2012, New York, NY, USA, May 19 - 22, 2012}, pages
  459--478, 2012{\natexlab{a}}.
\newblock \doi{10.1145/2213977.2214021}.
\newblock URL \url{https://doi.org/10.1145/2213977.2214021}.

\bibitem[Cai et~al.(2012{\natexlab{b}})Cai, Daskalakis, and Weinberg]{CaiDW12b}
Yang Cai, Constantinos Daskalakis, and S.~Matthew Weinberg.
\newblock Optimal multi-dimensional mechanism design: Reducing revenue to
  welfare maximization.
\newblock In \emph{53rd Annual {IEEE} Symposium on Foundations of Computer
  Science, {FOCS} 2012, New Brunswick, NJ, USA, October 20-23, 2012}, pages
  130--139, 2012{\natexlab{b}}.
\newblock \doi{10.1109/FOCS.2012.88}.
\newblock URL \url{https://doi.org/10.1109/FOCS.2012.88}.

\bibitem[Cai et~al.(2016)Cai, Devanur, and Weinberg]{CaiDW16}
Yang Cai, Nikhil Devanur, and S.~Matthew Weinberg.
\newblock A duality based unified approach to bayesian mechanism design.
\newblock In \emph{Proceedings of the 48th ACM Conference on Theory of
  Computation(STOC)}, 2016.

\bibitem[Cole and Roughgarden(2014)]{cole2014sample}
Richard Cole and Tim Roughgarden.
\newblock The sample complexity of revenue maximization.
\newblock In \emph{Proceedings of the forty-sixth annual ACM symposium on
  Theory of computing}, pages 243--252, 2014.

\bibitem[Conitzer and Sandholm(2002)]{conitzer2002complexity}
Vincent Conitzer and Tuomas Sandholm.
\newblock Complexity of mechanism design.
\newblock \emph{arXiv preprint cs/0205075}, 2002.

\bibitem[Conitzer and Sandholm(2004)]{conitzer2004self}
Vincent Conitzer and Tuomas Sandholm.
\newblock Self-interested automated mechanism design and implications for
  optimal combinatorial auctions.
\newblock In \emph{Proceedings of the 5th ACM conference on Electronic
  commerce}, pages 132--141, 2004.

\bibitem[Daskalakis and Weinberg(2012)]{daskalakis2012symmetries}
Constantinos Daskalakis and Seth~Matthew Weinberg.
\newblock Symmetries and optimal multi-dimensional mechanism design.
\newblock In \emph{Proceedings of the 13th ACM Conference on Electronic
  Commerce}, pages 370--387, 2012.

\bibitem[Daskalakis et~al.(2014)Daskalakis, Deckelbaum, and
  Tzamos]{DaskalakisDT14}
Constantinos Daskalakis, Alan Deckelbaum, and Christos Tzamos.
\newblock {The Complexity of Optimal Mechanism Design}.
\newblock In \emph{the 25th ACM-SIAM Symposium on Discrete Algorithms (SODA)},
  2014.

\bibitem[Daskalakis et~al.(2017)Daskalakis, Deckelbaum, and
  Tzamos]{daskalakis2017strong}
Constantinos Daskalakis, Alan Deckelbaum, and Christos Tzamos.
\newblock Strong duality for a multiple-good monopolist.
\newblock \emph{Econometrica}, 85\penalty0 (3):\penalty0 735--767, 2017.

\bibitem[Devanur et~al.(2016)Devanur, Huang, and Psomas]{DevanurHP16}
Nikhil~R. Devanur, Zhiyi Huang, and Christos{-}Alexandros Psomas.
\newblock The sample complexity of auctions with side information.
\newblock In \emph{Proceedings of the 48th Annual {ACM} {SIGACT} Symposium on
  Theory of Computing, {STOC} 2016, Cambridge, MA, USA, June 18-21, 2016},
  pages 426--439, 2016.
\newblock \doi{10.1145/2897518.2897553}.
\newblock URL \url{http://doi.acm.org/10.1145/2897518.2897553}.

\bibitem[Dhangwatnotai et~al.(2015)Dhangwatnotai, Roughgarden, and
  Yan]{DhangwatnotaiRY15}
Peerapong Dhangwatnotai, Tim Roughgarden, and Qiqi Yan.
\newblock Revenue maximization with a single sample.
\newblock \emph{Games and Economic Behavior}, 91:\penalty0 318--333, 2015.
\newblock \doi{10.1016/j.geb.2014.03.011}.
\newblock URL \url{https://doi.org/10.1016/j.geb.2014.03.011}.

\bibitem[Duetting et~al.(2019)Duetting, Feng, Narasimhan, Parkes, and
  Ravindranath]{dutting2017optimal}
Paul Duetting, Zhe Feng, Harikrishna Narasimhan, David Parkes, and Sai~Srivatsa
  Ravindranath.
\newblock Optimal auctions through deep learning.
\newblock In Kamalika Chaudhuri and Ruslan Salakhutdinov, editors,
  \emph{Proceedings of the 36th International Conference on Machine Learning},
  volume~97 of \emph{Proceedings of Machine Learning Research}, pages
  1706--1715, Long Beach, California, USA, 09--15 Jun 2019. PMLR.
\newblock URL \url{http://proceedings.mlr.press/v97/duetting19a.html}.

\bibitem[Dughmi et~al.(2014)Dughmi, Han, and Nisan]{dughmi2014sampling}
Shaddin Dughmi, Li~Han, and Noam Nisan.
\newblock Sampling and representation complexity of revenue maximization.
\newblock In \emph{International Conference on Web and Internet Economics},
  pages 277--291. Springer, 2014.

\bibitem[D{\"u}tting et~al.(2015)D{\"u}tting, Fischer, Jirapinyo, Lai, Lubin,
  and Parkes]{dutting2015payment}
Paul D{\"u}tting, Felix Fischer, Pichayut Jirapinyo, John~K Lai, Benjamin
  Lubin, and David~C Parkes.
\newblock Payment rules through discriminant-based classifiers.
\newblock \emph{ACM Transactions on Economics and Computation (TEAC)},
  3\penalty0 (1):\penalty0 1--41, 2015.

\bibitem[Feng et~al.(2018)Feng, Narasimhan, and Parkes]{feng2018deep}
Zhe Feng, Harikrishna Narasimhan, and David~C Parkes.
\newblock Deep learning for revenue-optimal auctions with budgets.
\newblock In \emph{Proceedings of the 17th International Conference on
  Autonomous Agents and Multiagent Systems}, pages 354--362. International
  Foundation for Autonomous Agents and Multiagent Systems, 2018.

\bibitem[Golowich et~al.(2018)Golowich, Narasimhan, and
  Parkes]{golowich2018deep}
Noah Golowich, Harikrishna Narasimhan, and David~C. Parkes.
\newblock Deep learning for multi-facility location mechanism design.
\newblock In \emph{Proceedings of the 17th International Joint Conference on
  Artificial Intelligence (IJCAI 2018)}, pages 261--267, 2018.
\newblock URL
  \url{https://econcs.seas.harvard.edu/files/econcs/files/golowich_ijcai18.pdf}.

\bibitem[Gonczarowski and Nisan(2017)]{GonczarowskiN17}
Yannai~A. Gonczarowski and Noam Nisan.
\newblock Efficient empirical revenue maximization in single-parameter auction
  environments.
\newblock In \emph{Proceedings of the 49th Annual {ACM} {SIGACT} Symposium on
  Theory of Computing, {STOC} 2017, Montreal, QC, Canada, June 19-23, 2017},
  pages 856--868, 2017.
\newblock \doi{10.1145/3055399.3055427}.
\newblock URL \url{http://doi.acm.org/10.1145/3055399.3055427}.

\bibitem[Gonczarowski and Weinberg(2018)]{GonczarowskiW18}
Yannai~A. Gonczarowski and S.~Matthew Weinberg.
\newblock The sample complexity of up-to-$\varepsilon$ multidimensional revenue
  maximization.
\newblock In \emph{59th {IEEE} Annual Symposium on Foundations of Computer
  Science, {FOCS}}, 2018.

\bibitem[Guo et~al.(2019)Guo, Huang, and Zhang]{GuoHZ19}
Chenghao Guo, Zhiyi Huang, and Xinzhi Zhang.
\newblock Settling the sample complexity of single-parameter revenue
  maximization.
\newblock In \emph{Proceedings of the 51st Annual {ACM} {SIGACT} Symposium on
  Theory of Computing, {STOC} 2019, Phoenix, AZ, USA, June 23-26, 2019.}, pages
  662--673, 2019.
\newblock \doi{10.1145/3313276.3316325}.
\newblock URL \url{https://doi.org/10.1145/3313276.3316325}.

\bibitem[Guo and Conitzer(2010)]{guo2010computationally}
Mingyu Guo and Vincent Conitzer.
\newblock Computationally feasible automated mechanism design: General approach
  and case studies.
\newblock In \emph{Twenty-Fourth AAAI Conference on Artificial Intelligence},
  2010.

\bibitem[Hart and Nisan(2012)]{HartN12}
Sergiu Hart and Noam Nisan.
\newblock {Approximate Revenue Maximization with Multiple Items}.
\newblock In \emph{the 13th ACM Conference on Electronic Commerce (EC)}, 2012.

\bibitem[Hart and Nisan(2013)]{HartN13}
Sergiu Hart and Noam Nisan.
\newblock The menu-size complexity of auctions.
\newblock In \emph{the 14th ACM Conference on Electronic Commerce (EC)}, 2013.

\bibitem[Hart and Reny(2015)]{HartR15}
Sergiu Hart and Philip~J. Reny.
\newblock {Maximizing Revenue with Multiple Goods: Nonmonotonicity and Other
  Observations}.
\newblock \emph{Theoretical Economics}, 10\penalty0 (3):\penalty0 893--922,
  2015.

\bibitem[Hartford et~al.(2018)Hartford, Graham, Leyton-Brown, and
  Ravanbakhsh]{hartford2018deep}
Jason Hartford, Devon~R Graham, Kevin Leyton-Brown, and Siamak Ravanbakhsh.
\newblock Deep models of interactions across sets.
\newblock \emph{arXiv preprint arXiv:1803.02879}, 2018.

\bibitem[Hartline and Roughgarden(2009)]{HartlineR09}
Jason~D. Hartline and Tim Roughgarden.
\newblock Simple versus optimal mechanisms.
\newblock In \emph{ACM Conference on Electronic Commerce}, pages 225--234,
  2009.

\bibitem[Hartline and Taggart(2019)]{HartlineT19}
Jason~D. Hartline and Samuel Taggart.
\newblock Sample complexity for non-truthful mechanisms.
\newblock In \emph{Proceedings of the 2019 {ACM} Conference on Economics and
  Computation, {EC} 2019, Phoenix, AZ, USA, June 24-28, 2019.}, pages 399--416,
  2019.
\newblock \doi{10.1145/3328526.3329632}.
\newblock URL \url{https://doi.org/10.1145/3328526.3329632}.

\bibitem[Huang et~al.(2018)Huang, Mansour, and Roughgarden]{huang2018making}
Zhiyi Huang, Yishay Mansour, and Tim Roughgarden.
\newblock Making the most of your samples.
\newblock \emph{SIAM Journal on Computing}, 47\penalty0 (3):\penalty0 651--674,
  2018.

\bibitem[Jin et~al.(2019{\natexlab{a}})Jin, Lu, Qi, Tang, and Xiao]{JinLQTX19}
Yaonan Jin, Pinyan Lu, Qi~Qi, Zhihao~Gavin Tang, and Tao Xiao.
\newblock Tight approximation ratio of anonymous pricing.
\newblock In Moses Charikar and Edith Cohen, editors, \emph{Proceedings of the
  51st Annual {ACM} {SIGACT} Symposium on Theory of Computing, {STOC} 2019,
  Phoenix, AZ, USA, June 23-26, 2019}, pages 674--685. {ACM},
  2019{\natexlab{a}}.
\newblock \doi{10.1145/3313276.3316331}.
\newblock URL \url{https://doi.org/10.1145/3313276.3316331}.

\bibitem[Jin et~al.(2019{\natexlab{b}})Jin, Lu, Tang, and Xiao]{JinLTX19}
Yaonan Jin, Pinyan Lu, Zhihao~Gavin Tang, and Tao Xiao.
\newblock Tight revenue gaps among simple mechanisms.
\newblock In \emph{Proceedings of the Thirtieth Annual ACM-SIAM Symposium on
  Discrete Algorithms}, pages 209--228. SIAM, 2019{\natexlab{b}}.

\bibitem[Kingma and Ba(2014)]{kingma2014adam}
Diederik~P Kingma and Jimmy Ba.
\newblock Adam: A method for stochastic optimization.
\newblock \emph{arXiv preprint arXiv:1412.6980}, 2014.

\bibitem[Lahaie(2011)]{lahaie2011kernel}
S{\'e}bastien Lahaie.
\newblock A kernel-based iterative combinatorial auction.
\newblock In \emph{Twenty-Fifth AAAI Conference on Artificial Intelligence},
  2011.

\bibitem[Li and Yao(2013)]{LiY13}
Xinye Li and Andrew Chi-Chih Yao.
\newblock On revenue maximization for selling multiple independently
  distributed items.
\newblock \emph{Proceedings of the National Academy of Sciences}, 110\penalty0
  (28):\penalty0 11232--11237, 2013.

\bibitem[Manelli and Vincent(2006)]{manelli2006bundling}
Alejandro Manelli and Daniel Vincent.
\newblock Bundling as an optimal selling mechanism for a multiple-good
  monopolist.
\newblock \emph{Journal of Economic Theory}, 127\penalty0 (1):\penalty0 1--35,
  2006.

\bibitem[Manelli and Vincent(2010)]{manelli2010bayesian}
Alejandro~M Manelli and Daniel~R Vincent.
\newblock Bayesian and dominant-strategy implementation in the independent
  private-values model.
\newblock \emph{Econometrica}, 78\penalty0 (6):\penalty0 1905--1938, 2010.

\bibitem[Maron et~al.(2018)Maron, Ben-Hamu, Shamir, and
  Lipman]{maron2018invariant}
Haggai Maron, Heli Ben-Hamu, Nadav Shamir, and Yaron Lipman.
\newblock Invariant and equivariant graph networks.
\newblock \emph{arXiv preprint arXiv:1812.09902}, 2018.

\bibitem[Medina and Mohri(2014)]{medina2014learning}
Andres~M Medina and Mehryar Mohri.
\newblock Learning theory and algorithms for revenue optimization in second
  price auctions with reserve.
\newblock In \emph{Proceedings of the 31st International Conference on Machine
  Learning (ICML-14)}, pages 262--270, 2014.

\bibitem[Morgenstern and Roughgarden(2016)]{morgenstern2016learning}
Jamie Morgenstern and Tim Roughgarden.
\newblock Learning simple auctions.
\newblock In \emph{Conference on Learning Theory}, pages 1298--1318, 2016.

\bibitem[Morgenstern and Roughgarden(2015)]{morgenstern2015pseudo}
Jamie~H Morgenstern and Tim Roughgarden.
\newblock On the pseudo-dimension of nearly optimal auctions.
\newblock In \emph{Advances in Neural Information Processing Systems}, pages
  136--144, 2015.

\bibitem[Myerson(1981)]{myerson1981optimal}
Roger~B Myerson.
\newblock Optimal auction design.
\newblock \emph{Mathematics of operations research}, 6\penalty0 (1):\penalty0
  58--73, 1981.

\bibitem[Narasimhan and Parkes(2016)]{narasimhan2016general}
Harikrishna Narasimhan and David~C Parkes.
\newblock A general statistical framework for designing strategy-proof
  assignment mechanisms.
\newblock In \emph{UAI'16 Proceedings of the Thirty-Second Conference on
  Uncertainty in Artificial Intelligence}, 2016.

\bibitem[Pavlov(2011)]{Pavlov11}
Gregory Pavlov.
\newblock Optimal mechanism for selling two goods.
\newblock \emph{The B.E. Journal of Theoretical Economics}, 11\penalty0 (3),
  2011.

\bibitem[Roughgarden and Schrijvers(2016)]{RoughgardenS16}
Tim Roughgarden and Okke Schrijvers.
\newblock Ironing in the dark.
\newblock In \emph{Proceedings of the 2016 {ACM} Conference on Economics and
  Computation, {EC} '16, Maastricht, The Netherlands, July 24-28, 2016}, pages
  1--18, 2016.
\newblock \doi{10.1145/2940716.2940723}.
\newblock URL \url{http://doi.acm.org/10.1145/2940716.2940723}.

\bibitem[Sandholm and Likhodedov(2015)]{sandholm2015automated}
Tuomas Sandholm and Anton Likhodedov.
\newblock Automated design of revenue-maximizing combinatorial auctions.
\newblock \emph{Operations Research}, 63\penalty0 (5):\penalty0 1000--1025,
  2015.

\bibitem[Shen et~al.(2019)Shen, Tang, and Zuo]{shen2019automated}
Weiran Shen, Pingzhong Tang, and Song Zuo.
\newblock Automated mechanism design via neural networks.
\newblock In \emph{Proceedings of the 18th International Conference on
  Autonomous Agents and Multiagent Systems}, pages 215--223. International
  Foundation for Autonomous Agents and Multiagent Systems, 2019.

\bibitem[Syrgkanis(2017)]{syrgkanis2017sample}
Vasilis Syrgkanis.
\newblock A sample complexity measure with applications to learning optimal
  auctions.
\newblock In \emph{Advances in Neural Information Processing Systems}, pages
  5352--5359, 2017.

\bibitem[Tacchetti et~al.(2019)Tacchetti, Strouse, Garnelo, Graepel, and
  Bachrach]{tacchetti2019neural}
Andrea Tacchetti, DJ~Strouse, Marta Garnelo, Thore Graepel, and Yoram Bachrach.
\newblock A neural architecture for designing truthful and efficient auctions.
\newblock \emph{arXiv preprint arXiv:1907.05181}, 2019.

\bibitem[Vickrey(1961)]{vickrey1961counterspeculation}
William Vickrey.
\newblock Counterspeculation, auctions, and competitive sealed tenders.
\newblock \emph{The Journal of finance}, 16\penalty0 (1):\penalty0 8--37, 1961.

\bibitem[Wang and Tang(2014)]{WangT14}
Zihe Wang and Pingzhong Tang.
\newblock Optimal mechanisms with simple menus.
\newblock In \emph{{ACM} Conference on Economics and Computation, {EC} '14,
  Stanford , CA, USA, June 8-12, 2014}, pages 227--240, 2014.
\newblock \doi{10.1145/2600057.2602863}.
\newblock URL \url{https://doi.org/10.1145/2600057.2602863}.

\bibitem[Zaheer et~al.(2017)Zaheer, Kottur, Ravanbakhsh, Poczos, Salakhutdinov,
  and Smola]{zaheer2017deep}
Manzil Zaheer, Satwik Kottur, Siamak Ravanbakhsh, Barnabas Poczos, Russ~R
  Salakhutdinov, and Alexander~J Smola.
\newblock Deep sets.
\newblock In \emph{Advances in neural information processing systems}, pages
  3391--3401, 2017.

\end{thebibliography}
\bibliographystyle{plainnat}

\onecolumn

\appendix

\appendix

\section{Permutation-equivariant network}\label{app:permeq_net}

In this section, we remind the  \textit{exchangeable matrix layers} introduced by \citet{hartford2018deep}. These layers are are a generalization of the deep sets architecture by  \citet{zaheer2017deep}. We briefly describe this architecture here and invite the reader to look at the original paper for details. \\

The architecture consists in several layers and each of them is constituted of multiple channels. Each layer is specified by the number of input channels $K$ channels and the number of outputs channels $O$. The input of such a layer is a tensor $B$ of size $(K,n,m)$ and the output is another tensor $Y$ of size  $(O,n,m)$.  The first element of these tensor is the channel number. In the following we will denote by $B^{(k)}_{i,j}$ the element of $B$ of index $(k,i,j)$ and similarly for  $Y^{(o)}_{i,j}$.\\

In addition to the $K$ and $O$, an exchangeable layer is defined by a set of five weights $w_1 , w_2, w_3, w_4 \in \mathbb{R}^{K \times O}$ and $w_5 \in \mathbb{R}$. Given these weights, the element $(i,j)$ of the $o$-th output channel $Y_{i,j}^{(o)}$ is given by:
\begin{equation}\label{eq:mult_channel}
\small
\begin{aligned}
\small
 Y_{i,j}^{(o)}=\sigma \Bigg ( &\sum_{k=1}^K w_1^{(k,o)}B_{i,j}^{(k)}+\frac{w_2^{(k,o)}}{n}\sum_{i'}B_{i',j}^{(k)} \\
 &+\frac{w_3^{(k,o)}}{m}\sum_{j'}B_{i,j'}^{(k)}       +\frac{w_4^{(k,o)}}{nm}\sum_{i',j'}B_{i',j'}^{(k)}+w_5^{(o)}
 \Bigg)
 \end{aligned}
\end{equation}
This layer preserved permutation-equivariance. This was first proven in \citet{hartford2018deep}.  Additionally, the number of parameters of each layer only depends on $K$ and $O$ is does not depend on the dimension of the input (i.e. $m$ and $n$). In particular, we can apply this exchangeable layer to any tensor of size $(K,n',m')$ for any value of $n'$ and $m'$ and the resulting output will be a tensor of size $(O,n',m')$.\\

We can compose these exchangeable layers as long as the number of channels of the output of one layer is equal to the number of input channels required by the following layer. By composing many such layer of this form we get a deep exchangeable neural network. This deep network preserved permutation-equivariance since this property is preserved by every layer. In addition, this network can be evaluated on an input of any dimension $n$ and $m$. We use this property of the network to test our mechanisms on settings with different number of bidders and objects. Without this property out of setting generalization not possible.

\section{Proof of \autoref{prop:equiv_sol}}\label{app:thm}
Notation: For a matrix $B \in \mathbb{R}^{nm}$ we will denote the $i$th line by $B_i \in \mathbb{R}^{m} $ or $[B]_i \in \mathbb{R}^{m}$.Let $D$ denote an equivariant distribution on $\mathbb{R}^{nm}$ .  Let $g\colon\mathbb{R}^{nm} \rightarrow \mathbb{R}^{nm}$ and $p\colon\mathbb{R}^{nm}\rightarrow \mathbb{R}^{n} $  be solutions to the following problem: 
$$p = \mbox{argmax}\,\, \mathbb{E}_{B \sim D} \left[ \sum_{i=1}^n p_i(B) \right]$$
subject to:
$$\langle\, [g(B)]_i \, , \,B_i \, \rangle \geq p_i(B),$$
and 
$$\langle  \,[g(B_i,B_{-i})]_i \,, \,B_i \,\rangle -  p_i(B_i,B_{-i}) \geq \langle \,[g(B'_i,B_{-i})]_i \,,\,B_i \,\rangle -  p_i(B'_i,B_{-i}), \,\, \,\, \forall B'_i \in \mathbb{R}^m .$$

Let $\Pi_n$ and $\Pi_m$ be two permutation matrices of sizes $n$ and $m$. In particular $\Pi_n$ and $\Pi_m$ are orthogonal matrices and in the following we use that $\Pi_n^{-1} =\Pi_n^{T} $ and $\Pi_m^{-1} =\Pi_m^{T} $. Let's define: 

\begin{equation*}
\begin{aligned}
g^{\Pi_{n},\Pi_{m}}(B) &= \Pi_{n}^{-1}\, g(\Pi_{n}\,B\, \Pi_{m}) \, \Pi_{m}^{-1} \\
p^{\Pi_{n},\Pi_{m}}(B) &= \Pi_{n}^{-1} \, p(\Pi_{n}\,B\, \Pi_{m}).
\end{aligned}
\end{equation*}

Let's prove that if $(g,p)$ is a solution to the problem then so is  $(g^{\Pi_{n},\Pi_{m}},p^{\Pi_{n},\Pi_{m}})$. First we show that $(g^{\Pi_{n},\Pi_{m}},p^{\Pi_{n},\Pi_{m}})$ still satisfy the previous constraints.
\begin{equation*}
\begin{aligned}
\langle [g^{\Pi_{n},\Pi_{m}}(B)]_i \, , \, B_i \, \rangle & = \langle [\Pi_{n}^{-1}g(\Pi_{n}\,B\,\Pi_{m})\Pi_{m}^{-1}]_i \, , \, B_i \, \rangle \\
& = \langle [\Pi_{n}^{-1}g(\Pi_{n}\,B\,\Pi_{m})]_i \Pi_{m}^{-1} \, , \, B_i \, \rangle \\
& = \langle [\Pi_{n}^{-1}g(\Pi_{n}\,B\,\Pi_{m})]_i \, , \, B_i \Pi_{m} \, \rangle \\
& = \langle [\Pi_{n}^{-1}g(\Pi_{n}\,B\,\Pi_{m})]_i \, , \, [B \Pi_{m}]_i \, \rangle \\
& = \langle [\Pi_{n}^{-1}g(\Pi_{n}\,B\,\Pi_{m})]_i \, , \, [B \Pi_{m}]_i \, \rangle.
\end{aligned}
\end{equation*}
Let's denote by $\phi$ the permutation on the indices corresponding to the $\Pi_{n}$ permutation. then we have: 

\begin{equation*}
\begin{aligned}
[\Pi_{n}^{-1}g(\Pi_{n}\,B\,\Pi_{m})]_i  &= [g(\Pi_{n}\,B\,\Pi_{m})]_{\phi^{-1}(i)}\\
[B \Pi_{m}]_i &= [\Pi_{n}B \Pi_{m}]_{\phi^{-1}(i)}.
\end{aligned}
\end{equation*}
This gives us that: 

\begin{equation*}
\begin{aligned}
\langle [g^{\Pi_{n},\Pi_{m}}(B)]_i \, , \, B_i \, \rangle & = \langle [\Pi_{n}^{-1}g(\Pi_{n}\,B\,\Pi_{m})]_i \, , \, [B \Pi_{m}]_i \, \rangle \\
& = \langle [g(\Pi_{n}\,B\,\Pi_{m})]_{\phi^{-1}(i)} \, , \, [\Pi_{n}B \Pi_{m}]_{\phi^{-1}(i)} \, \rangle \\ 
& \geq  [p(\Pi_{n}B \Pi_{m})]_{\phi^{-1}(i)}\\
& = [\Pi_{n}^{-1}p(\Pi_{n}B \Pi_{m})]_{i} \\
&= [p^{\Pi_{n},\Pi_{m}}(B)]_{i}.
\end{aligned}
\end{equation*}
This shows that $(g^{\Pi_{n},\Pi_{m}},p^{\Pi_{n},\Pi_{m}})$ satisfies the first constraint. We now move to the second constraint. \\

Let's write $\tilde{B} = (B'_{i},B_{-i})$. As a reminder, this is the matrix $B$ where the $i$th line has been replaced with $B'_i$. We need to show that: 

$$\langle  \,[g^{\Pi_{n},\Pi_{m}}(B)]_i \,, \,B_i \,\rangle -  p_i^{\Pi_{n},\Pi_{m}}(B) \geq \langle \,[g^{\Pi_{n},\Pi_{m}}(\tilde{B})]_i \,,\,B_i \,\rangle -  p^{\Pi_{n},\Pi_{m}}_i(\tilde{B}).$$

Using the previous computations we find that: 
$$\langle  \,[g^{\Pi_{n},\Pi_{m}}(B)]_i \,, \,B_i \,\rangle -  p_i^{\Pi_{n},\Pi_{m}}(B)  = \langle [g(\Pi_{n}\,B\,\Pi_{m})]_{\phi^{-1}(i)} \, , \, [\Pi_{n}B \Pi_{m}]_{\phi^{-1}(i)} \, \rangle - [p(\Pi_{n}B \Pi_{m})]_{\phi^{-1}(i)}, $$
where $\phi$ is the permutation associated with $\Pi_n$. Since $g$ and $p$ satisfy the second constraint we have:  
\begin{equation*}
\begin{aligned}    
\langle  \,[g^{\Pi_{n},\Pi_{m}}(B)]_i \,, \,B_i \,\rangle -  p_i^{\Pi_{n},\Pi_{m}}(B) & = \langle [g(\Pi_{n}\,B\,\Pi_{m})]_{\phi^{-1}(i)} \, , \, [\Pi_{n}B \Pi_{m}]_{\phi^{-1}(i)} \, \rangle - [p(\Pi_{n}B \Pi_{m})]_{\phi^{-1}(i)}\\
 &\geq  \langle [g(\Pi_{n}\,\tilde{B}\,\Pi_{m})]_{\phi^{-1}(i)} \, , \, [\Pi_{n}\tilde{B} \Pi_{m}]_{\phi^{-1}(i)} \, \rangle - [p(\Pi_{n}\tilde{B} \Pi_{m})]_{\phi^{-1}(i)}
 \\
 &= \langle  \,[g^{\Pi_{n},\Pi_{m}}(\tilde{B})]_i \,, \,B_i \,\rangle -  p_i^{\Pi_{n},\Pi_{m}}(\tilde{B}).
\end{aligned}  
\end{equation*}

This concludes the proof that $(g^{\Pi_{n},\Pi_{m}},p^{\Pi_{n},\Pi_{m}})$ satisfy the constraints. Now we have to show that $p^{\Pi_{n},\Pi_{m}}$ is optimal.

\begin{equation*}
\begin{aligned}
\mathbb{E}_{B \sim D} \left[ \sum_{i=1}^n p^{\Pi_{n},\Pi_{m}}(B) \right] &=  \mathbb{E}_{B \sim D} \left[ \langle \,  p^{\Pi_{n},\Pi_{m}}(B) \, , \bf{1} \, \rangle \right]\\
&=  \mathbb{E}_{B \sim D} \left[ \langle \,   \Pi_{n}^{-1} \, p(\Pi_{n}\,B\, \Pi_{m}) \, , \bf{1} \, \rangle \right] \\
&=  \mathbb{E}_{B \sim D} \left[ \langle \,   p(\Pi_{n}\,B\, \Pi_{m}) \, , \bf{1} \, \rangle \right]\\
&=  \mathbb{E}_{B \sim D} \left[ \langle \,   p(B) \, , \bf{1} \, \rangle \right]\\
&= \mathbb{E}_{B \sim D} \left[ \sum_{i=1}^n p_i(B) \right],
\end{aligned}
\end{equation*}
where we used that $\Pi_n^{-1} =\Pi_n^{T} $,  $ \Pi_n \bf{1} = \bf{1} $  and that $\Pi_{n}\,B\, \Pi_{m} \sim D $ since $D$ is an equivariant distribution. This shows that if $p$ is optimal then $p^{\Pi_{n},\Pi_{m}}$ is also optimal since they have the same expectation. We conclude that  $(g^{\Pi_{n},\Pi_{m}},p^{\Pi_{n},\Pi_{m}})$ is an optimal solution.  Let's define
\begin{equation*}
\begin{aligned}
\tilde{g}(B) &= \mathbb{E}_{\Pi_{n},\Pi_{m}} \left[ g^{\Pi_{n},\Pi_{m}}(B)\right] \\
\tilde{p}(B) &= \mathbb{E}_{\Pi_{n},\Pi_{m}} \left[ p^{\Pi_{n},\Pi_{m}}(B)\right].
\end{aligned}
\end{equation*}

Here, in the expectation, $\Pi_{n}$ and $\Pi_{m}$ are drawn uniformly at random. Since the problem and constraints are convex, $(\tilde{g},\tilde{p})$ is also an optimal solution to the problem as a convex combination of optimal solutions. Let's prove that $\tilde{g}$ and $\tilde{p}$ are equivariant functions.

\begin{equation*}
\begin{aligned}
\tilde{g}(\Pi_{n}\,B\,\Pi_{m}) &= \mathbb{E}_{\Pi'_{n},\Pi'_{m}} \left[ g^{\Pi'_{n},\Pi'_{m}}(\Pi_{n}\,B\,\Pi_{m})\right] \\ 
&= \mathbb{E}_{\Pi'_{n},\Pi'_{m}} \left[ {\Pi'_{n}}^{-1}g(\Pi'_{m}\,\Pi_{n}\,B\,\Pi_{m}\,\Pi'_{m}){\Pi'_{m}}^{-1}\right] \\ 
&= {\Pi_{n}}^{-1} \,\, \mathbb{E}_{\Pi'_{n},\Pi'_{m}} \left[ {(\Pi'_{n}\Pi_{n})}^{-1}g(\Pi'_{n}\,\Pi_{n}\,B\,\Pi_{m}\,\Pi'_{m}){(\Pi_{m}\Pi'_{m})}^{-1}\right]\,\, {\Pi_{m}}^{-1}. \\ 
\end{aligned}
\end{equation*}
If ${\Pi'_{n}}$ and ${\Pi'_{m}}$ are uniform among permutation then so is $\Pi'_{n}\Pi_{n}$ and $\Pi'_{m}\Pi_{m}$. So through a change of variable we find that: 

\begin{equation*}
\begin{aligned}
\tilde{g}(\Pi_{n}\,B\,\Pi_{m}) 
&= {\Pi_{n}}^{-1} \,\, \mathbb{E}_{\Pi'_{n},\Pi'_{m}} \left[ {\Pi'_{n}}^{-1}g(\Pi'_{n}\,B\,\Pi'_{m}){\Pi'_{m}}^{-1}\right]\,\, {\Pi_{m}}^{-1} \\ 
&= {\Pi_{n}}^{-1} \,\, \tilde{g}(B) \,\, {\Pi_{m}}^{-1}.
\end{aligned}
\end{equation*}
This shows that $\tilde{g}$ is equivariant. The proof that $\tilde{p}$ is equivariant is similar. 

\begin{equation*}
\begin{aligned}
\tilde{p}(\Pi_{n}\,B\,\Pi_{m}) &= \mathbb{E}_{\Pi'_{n},\Pi'_{m}} \left[ p^{\Pi'_{n},\Pi'_{m}}(\Pi_{n}\,B\,\Pi_{m})\right] \\ 
&= \mathbb{E}_{\Pi'_{n},\Pi'_{m}} \left[ {\Pi'_{n}}^{-1}p(\Pi'_{m}\,\Pi_{n}\,B\,\Pi_{m}\,\Pi'_{m})\right] \\ 
&= {\Pi_{n}}^{-1} \,\, \mathbb{E}_{\Pi'_{n},\Pi'_{m}} \left[ {(\Pi'_{n}\Pi_{n})}^{-1}p(\Pi'_{n}\,\Pi_{n}\,B\,\Pi_{m}\,\Pi'_{m})\right]. \\ 
\end{aligned}
\end{equation*}
By doing a change of variable as before we find: 
\begin{equation*}
\begin{aligned}
\tilde{p}(\Pi_{n}\,B\,\Pi_{m}) 
&= {\Pi_{n}}^{-1} \,\, \mathbb{E}_{\Pi'_{n},\Pi'_{m}} \left[ {\Pi'_{n}}^{-1}p(\Pi'_{n}\,B\,\Pi'_{m})\right] \\ 
&= {\Pi_{n}}^{-1} \,\, \tilde{p}(B),
\end{aligned}
\end{equation*}
$(\tilde{g},\tilde{p})$ is an equivariant optimal solution, this concludes the proof.

\section{Optimization and training procedures}\label{app:opt_train}

\begin{algorithm}[t]
\caption{Training Algorithm} 
\label{alg:langevin_dynamics}
\begin{algorithmic}[1]
\State \textbf{Input}: Minibatches $\mathcal{S}_1,\dots,\mathcal{S}_T$ of size $B$
\State \textbf{Parameters}: $\gamma >0, \, \eta >0, \, c>0, \, R\in \mathbb{N}, \,  T\in \mathbb{N}, \,T_{\rho}\in \mathbb{N}, \,T_{\lambda}\in \mathbb{N}.$
\State \textbf{Initialize Parameters}: $ \rho^0 \in \mathbb{R}, \, w^0\in \mathbb{R}^d, \, \lambda^0\in\mathbb{R}^n,  $

\State \textbf{Initialize Misreports:} ${v_i'}^{(\ell)}\in V_i, \,\, \forall \ell \in [B], \, i\in N.$ \\
\For {$t=0,\dots,T$}
        \State Receive minibatch $\mathcal{S}_t = \{V^{(1)},\dots,V^{(B)}\}.$
        \For{$r=0,\dots,R$}
        \State
        \vspace{-1.1cm}
        
        \begin{align*}
            \hspace{.6cm}&\forall \ell \in [B], \, i\in n:\\
            &{v_i'}^{(\ell)} \leftarrow {v_i'}^{(\ell)}+\gamma \nabla_{v_i'}u_i^{w_t} ({v_i}^{(\ell)};({v_i'}^{(\ell)},V_{-i}^{(\ell)}))
        \end{align*}
        \EndFor
        
        \vspace{-.4cm}
        \\
        \State Get Lagrangian gradient using \eqref{eq:grad_Lag} and update $w^t$:\\
         \hspace{1cm} $w^{t+1}\leftarrow w^t-\eta \nabla_w \mathcal{L}_{\rho^t}(w^t). $
         \\
        \State Update $\rho$ once in $T_{\rho}$ iterations:
        \If{$t$ is a multiple of $T_{\rho}$}
        \State $\rho^{t+1} \leftarrow \rho^{t} + c $ 
        \Else
        \State $\rho^{t+1}\leftarrow \rho^t$
        \EndIf
         \\
        \State Update Lagrange multipliers once in $T_{\lambda}$ iterations:
        \If{$t$ is a multiple of $T_{\lambda}$}
        \State $\lambda_i^{t+1}\leftarrow \lambda_i^t + \rho^t \, \widehat{rgt}_i(w^{t}),\forall i\in N$
        \Else
        \State $\lambda^{t+1}\leftarrow \lambda^t$
        \EndIf
\EndFor
\end{algorithmic}
\end{algorithm}

Our training algorithm is the same as the one found in \cite{dutting2017optimal}. We made that choice to better illustrate the intrinsic advantages of our permutation equivariant architecture.  We include implementation details here for completeness and additional details can be found in the original paper. \\

We generate a training dataset of valuation profiles $\mathcal{S}$ that we then divide into mini-batches of size $B$. Typical sizes for $\mathcal{S}$ are $\{5000,50000,500000 \}$ and typical batch sizes are $\{50,500,50000\}$.  We train our networks over for several epochs (typically $\{50,80 \}$) and we apply a random shuffling of the training data for each new epoch. We denote the minibatch received at iteration $t$ by $\mathcal{S}_t=\{V^{(1)},\dots,V^{(B)}\}.$ The update on model parameters involves an unconstrained optimization of $\mathcal{L}_{\rho}$ over $w$ and is performed using a gradient-based optimizer. Let $\widehat{rgt}_i(w)$ be the empirical regret in \eqref{eq:emp_reg} computed on mini-batch $\mathcal{S}_t.$ The gradient of $\mathcal{L}_{\rho}$ with respect to $w$ is given by: 
\vspace*{-.27cm}
\begin{equation}\label{eq:grad_Lag}
\begin{split}
    \nabla_w \mathcal{L}_{\rho}(w)&=-\frac{1}{B}\sum_{\ell=1}^B\sum_{i \in N} \nabla_w p_i^w(V^{(\ell)})\\
    &\hspace{-.5cm}+\sum_{i \in N}\sum_{\ell=1}^B\lambda_i^t g_{\ell,i}+\rho_t \sum_{i \in N}\sum_{\ell=1}^B\widehat{rgt}_i(w)g_{\ell,i}, 
\end{split}
\end{equation}
where 
\begin{align*}
    g_{\ell,i}=\nabla_w\left[\max_{v_i'\in V_i} u_i^w(v_i^{(\ell)};(v_i',V_{-i}^{(\ell)}))-u_i^w(v_i^{(\ell)};(v_i^{(\ell)},V_{-i}^{(\ell)}))\right].
\end{align*}
The terms $\widehat{rgt}_i$ and $g_{\ell,i}$ requires us to compute the maximum over misreports for each bidder $i$ and valuation profile $\ell$. To compute this maximum we optimize the function $v_i' \to u_i^w(v_i^{(\ell)};(v_i',V_{-i}^{(\ell)}))$ using another gradient based optimizer. \\

For each $i$ and valuation profile $\ell$, we maintain a misreports valuation ${v_i'}^{(\ell)}$.  For every update on the model parameters $w^t,$  we perform $R$ gradient updates to compute the optimal misreports: ${v_i'}^{(\ell)}={v_i'}^{(\ell)}+\gamma \nabla_{{v_i'}^{(\ell)}}u_i^w(v_i^{(\ell)};({v_i'}^{(\ell)},V_{-i}^{(\ell)})),$ for some $\gamma >0.$  In our experiments, we use the Adam optimizer \citep{kingma2014adam} for updates on model $w$ and ${v_i'}^{(\ell)}. $ Typical values are $R = 25$ and $\gamma = 0.001$ for the training phase.  During testing, we use a larger number of step sizes $R_{test}$ to compute these optimal misreports and we try bigger number initialization, $N_{init}$, that are drawn from the same distribution of the valuations. Typical values are $R_{test} = \{200,300\}$ and $N_{init} = \{100,300\}$.
When the valuations are constrained to an interval (for instance $[0,1]$), this optimization inner loop becomes constrained and we make sure that the values we get for  $v_i'$ are realistic by projecting them to their domain after each gradient step. \\

The parameters $\lambda^t$ and $\rho_t$ in the Lagrangian are not constant but they are updated over time.  $\rho_t$ is initialized at a value $\rho_0$ is incremented every $T_{\rho}$ iterations, $\rho_{t+1} \leftarrow \rho_{t} + c $. Typical values are $\rho_{0} = \{0.25,1\}$,  $c = \{0.25,1, 5\}$ and $T_{\rho} = \{2, 5\}$ epochs. $\lambda_t$ is initialized at a value $\lambda_0$ is updates every $T_{\lambda}$ iterations according to $\lambda_i^{t+1} \leftarrow \lambda_i^t + \rho_t \, \widehat{rgt}_i(w^{t}),\forall i\in N$.  Typical values are $\lambda^{0}_i = \{0.25,1,5\}$ and $T_{\lambda} = \{2\}$ iterations.

\section{Setup} \label{app:setup}
 
We implemented our experiments using PyTorch. A typical deep exchangeable network consists of 3 hidden layers of 25 channels each. Depending on the experiment, we generated a dataset of $\{5000,50000,500000\}$ valuation profiles and chose mini batches of sizes $\{50,500,5000\}$ for training. 
The optimization of the augmented Lagrangian was typically run for $\{50, 80\}$ epochs. The value of $\rho$ in the augmented Lagrangian was set to $1.0$ and incremented every $2$ epochs. An update on $w^t$ was performed for every mini-batch using the Adam optimizer with a learning rate of $0.001$. For each update $w^t$, we ran $R=25$ misreport update steps with a learning rate of $0.001.$ An update on $\lambda^t$ was performed once every $100$ minibatches.

\end{document}